\documentclass[twocolumn]{aastex631}

\usepackage{amsmath}
\usepackage{multirow}
\usepackage{xcolor}
\usepackage{hyperref}
\graphicspath{{../../Figures/}}

\usepackage{color}

\newcommand{\kms}{\mathrm{km\,s^{-1}}}
\newcommand{\msun}{\mathrm{M_\odot}}
\newcommand{\rsun}{\mathrm{R_\odot}}

\shorttitle{A concordance model of NS kicks}
\shortauthors{Shariat et al.}

\begin{document}

\title{\Large A concordance model of neutron star kicks}

\author[0000-0002-4424-1840]{Cheyanne Shariat}
\affiliation{Department of Astronomy, California Institute of Technology, 1200 E. California Blvd, Pasadena, CA, 91125, USA}

\author[0000-0002-6871-1752]{Kareem El-Badry}
\affiliation{Department of Astronomy, California Institute of Technology, 1200 E. California Blvd, Pasadena, CA, 91125, USA}

\author[0000-0002-9802-9279]{Smadar Naoz}
\affiliation{Department of Physics and Astronomy, University of California, Los Angeles, Los Angeles, CA 90095, USA}
\affiliation{Mani L. Bhaumik Institute for Theoretical Physics, Department of Physics and Astronomy, UCLA, Los Angeles, CA 90095, USA}

\begin{abstract}
Neutron stars (NSs) receive natal kicks during supernovae that affect their survival in binaries and motion through the Galaxy. The global distribution of these kick velocities across the NS population remains uncertain. Here, we combine the orbital properties and birth rates of Gaia NS and Be+NS binaries with velocity constraints from young isolated pulsars to construct a concordance model of NS natal kicks. We find that all three populations can be jointly described by a shared bimodal kick distribution. The low-velocity component follows a lognormal distribution with $\mu_{\ln v}=1.87^{+0.65}_{-0.53}$ and $\sigma_{\ln v}=0.55^{+0.85}_{-0.44}$, corresponding to a median speed of $6.5^{+5.9}_{-2.7}~{\rm km~s^{-1}}$. The high-velocity component is consistent with the distribution inferred from young pulsars, which peaks near $200~{\rm km~s^{-1}}$. The low-velocity component contributes only $\sim10\%$ of all NS births, but produces most Gaia NS binaries and roughly half of Be+NS binaries. The same model predicts systemic recoil velocities consistent with the observed kinematics of Gaia NSs, Be+NS binaries, and NS low-mass X-ray binaries. For Gaia NSs, whose formation history remains unknown, we use this kick distribution to constrain their progenitors. We favor pre-supernova binaries containing intermediate-mass stripped stars with $\sim1~{\rm M_\odot}$ companions in au-scale orbits, and predict $\sim300$ such binaries in the Milky Way today, $\sim40$ in the LMC, and $\sim30$ in the SMC. Finally, we consider the implications of these kicks for NSs in hierarchical triples. We find that $1$--$7\%$ of Gaia NS triples and $2$--$20\%$ of Be+NS triples retain their tertiary after the supernova. Overall, our results suggest that a shared natal kick law can account for isolated pulsars and surviving NS binaries, where most NSs receive large kicks and low kicks are common among surviving binaries.
\end{abstract}


\section{Introduction}\label{sec:introduction}

Nearly all neutron star (NS) progenitors are born with companions.  Massive OB stars have binary or higher-order multiplicity fractions of $80$--$100\%$ \citep{Sana2012,Moe2017}, yet most young Galactic NSs are observed as isolated pulsars \citep{Lorimer2008}.  This suggests that a large fraction of massive-star binaries either merge before core collapse or become unbound during the supernova \citep[e.g.,][]{Brandt1995,Tauris1998,Renzo2019}.  The impulse delivered to the compact object during a supernova (the `natal kick') is a key ingredient for modeling the evolution of massive binaries and their downstream outcomes, including stripped-envelope supernovae, X-ray binaries, recycled pulsars, kilonovae, and gravitational-wave mergers \citep[e.g.,][]{Bhattacharya1991,Podsiadlowski2002,Tauris2012,Tauris2017,VignaGomez2018,Abbott2017}.

Although natal kicks are thought to arise during the supernova (SN), the specific mechanisms and subsequent kick magnitudes remain uncertain.
Large kicks are often attributed to hydrodynamic recoil from asymmetric ejecta, with contributions from anisotropic neutrino emission \citep[see][]{Burrows1995,Arras1999,Janka2017, Lai2004,Tamborra2014, Nagakura2019, Coleman2022, Burrows2024,Burrows2024a, JankaKresse2024}.  
Small kicks are expected from rapid, symmetric, or low-ejecta-mass explosions, such as electron-capture and ultra-stripped supernovae \citep[e.g.,][]{Nomoto1984,Nomoto1987,Podsiadlowski2004,Tauris2013,Tauris2015,Janka2017,Gessner2018,Tauris2017}.
Isolated pulsars show evidence for relatively large kicks, with inferred kicks ranging from $\sim50~\kms$ to $>1000~\kms$ \citep{Lyne1994,Hobbs2005,FaucherGiguere2006,Verbunt2017,Igoshev2020,Disberg2025}, although the shape of this distribution is still debated \citep[e.g.,][]{Arzoumanian2002,Verbunt2017,Igoshev2020,Disberg2025a,Disberg2025}. 
These measurements anchor the kick distribution of isolated NSs, but they do not directly probe the binaries that survive core collapse.

Binaries that remain bound after NS formation provide a complementary perspective.  Their existence already selects against very large kicks, and their post-SN orbits set the future evolution of the binary. 
One way to constrain natal kicks is by measuring the present-day motion of the surviving binaries through the Galaxy.  This approach has been applied to a variety of binaries containing NSs and black holes \citep[e.g.,][]{Repetto2012,Mandel2016,Atri2019,Fortin2022,ODoherty2023, Zhao2023,Disberg2024,Nagarajan2025a}.  The systemic velocity of these systems encodes the center-of-mass recoil imparted on the binary due to the kick, but does not explicitly isolate the natal kick that the compact object received.  Instead, it measures a combination of the natal kick on the compact object, acceleration due to symmetric mass loss, the pre-SN orbital velocity, and later evolution in the Galactic potential \citep{Blaauw1961,Hills1983,Kalogera1996,Kalogera1998,Atri2019,ODoherty2023,Disberg2024, Disberg2024a}.  This makes systemic velocities a useful but indirect constraint on the natal kick, given the uncertainties in Galactic phase, disk heating, ages, and pre-SN binary evolution, which can all blur the connection to the natal kick \citep{Wielen1977,Atri2019,Zhao2023,ODoherty2023}.

A complementary constraint on natal kicks can be inferred from the orbits of surviving binaries, particularly those whose orbits have changed little since the SN.
In that case, the present-day period and eccentricity can be modeled assuming an instantaneous explosion\footnote{An alternative is the ``rocket''-kick scenario, where delayed acceleration of the newborn NS modifies the post-SN orbit \citep[e.g.,][for applications to Gaia NSs]{Hirai2024, Baibhav2026}.} to infer the natal kick and mass loss.  This method allows inferences on the kicks received in wide Be+NS binaries and double NSs \citep[e.g.,][]{Beniamini2016,Andrews2019,Valli2025}.  
Wide Be+NS systems are especially useful because 
the pre-SN evolution is comparatively well understood, 
there has only been one kick since then, tides are inefficient over the Be-star lifetime for orbital periods $P_{\rm orb} \gtrsim5$~days, and pulse timing gives precise orbital periods and eccentricities \citep{Reig2011,Bashi2023}.  
\citet{Valli2025} used a sample of 23 Be-XRBs to infer two kick families: an ultralow-kick component with $v_{\rm kick}<10~\kms$ for the low-eccentricity systems and a $\sim100~\kms$ component for the high-eccentricity sequence, with the latter favoring polar alignment within a few degrees of the progenitor spin axis. 

The recently discovered Gaia NS candidates extend orbit-based kick constraints to a new regime.  The $27$ systems identified in {\it Gaia} DR3 contain solar-type stars with dark companions in the NS mass range\footnote{This selection criterion allows for contamination by ultramassive WDs or compact WD+WD binaries (see Section~\ref{subsubsec:contamination} for a discussion).}, on detached au-scale orbits \citep{ElBadry2024b,ElBadry2024d,ElBadry2026}.  Their observed periods are $P_{\rm orb}=189$--$1405$ d and their eccentricities are $e=0.12$--$0.79$.  At these separations, tides, winds, gravitational radiation, and later stellar evolution should not have changed the orbit discernibly since NS formation. Unlike double NSs, the interpretation involves one supernova rather than two, and unlike interacting X-ray binaries, the systems have not undergone Roche-lobe overflow after NS formation.  The Gaia NSs therefore provide one of the cleanest laboratories for constraining NS natal kicks.

The Gaia NSs also offer two key advantages that make them unique among most other NS binary populations.  First, they are nearby ($d\lesssim2~{\rm kpc}$) and have precise astrometry with extensive radial velocity follow-up, providing 6D kinematics for each binary \citep{ElBadry2024b,ElBadry2026}.  Second, the {\it Gaia} astrometric binary selection function is well characterized \citep{Halbwachs2023,ElBadry2024,Lam2025}, which is not the case for NS-LMXBs, for example \citep[e.g.,][]{Maccarone2022}.
Leveraging this, \citet{Shariat2026GaiaNSDemographics} modeled the {\it Gaia} selection function and used hierarchical Bayesian inference to constrain the intrinsic demographics of the Gaia NS population, including the distributions of period, eccentricities, masses, and space densities.
The resulting intrinsic population 
provides a benchmark for Gaia NS formation and kick models, where a successful model should match both the Gaia NS intrinsic period--eccentricity distribution and relative abundance in the Galaxy.

In this paper, we use the intrinsic Gaia NS population to infer the progenitor masses, mass loss, and natal kicks required to form them. We then connect our results with the broader NS population to provide a global view of NS kicks.
The remainder of the paper is structured as follows.  Section~\ref{sec:methods} describes the methodology.  Section~\ref{sec:kick_constraints} presents the Gaia NS kick constraints.  Section~\ref{sec:joint} combines Gaia NSs with Be+NS binaries to infer a shared parent kick law and its contribution to the full NS birth population.  Section~\ref{sec:kinematics} compares the orbital kick constraints to Gaia NS systemic velocities and NS-LMXB systemic velocities.  Section~\ref{sec:tertiaries} tests the survival of wide tertiary companions to Gaia NSs.  Section~\ref{sec:discussion} places the result in the context of Gaia NS progenitors, isolated pulsars, X-ray binaries, and physical implications for low supernova kicks. Section~\ref{sec:conclusions} summarizes the conclusions. Additional details are presented in Appendices
\ref{app:sn_formalism}, \ref{app:survivor_distributions}, \ref{app:presn_eccentricity}, \ref{app:toomre}, \ref{app:high_mhe}, and \ref{app:lowq_progenitor}. 

\section{Methodology}\label{sec:methods}

\begin{figure*}
    \centering
    \includegraphics[width=0.99\textwidth]{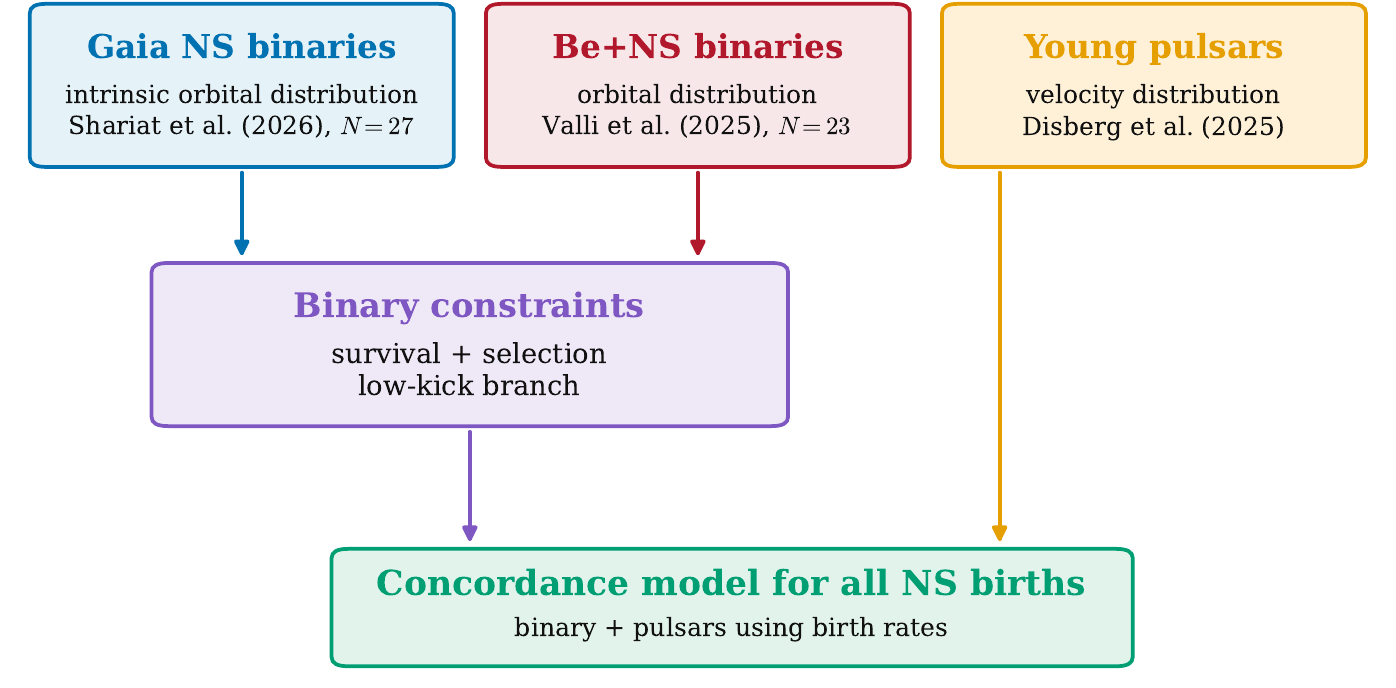}
    \caption{Overview of the natal kick inference. Gaia NS and Be+NS orbits constrain the kick law for binaries that survive the supernova. We combine this binary constraint with the young-pulsar kick law using their relative birth rates to infer the kick distribution for all NS births.}
    \label{fig:kick_inference_flowchart}
\end{figure*}

Figure~\ref{fig:kick_inference_flowchart} summarizes our approach. 
We first use the \emph{intrinsic} Gaia NS period--eccentricity distribution, birth rate, and kinematics to infer the pre-SN orbits, mass loss, and natal kicks required to form the population. Section~\ref{subsec:gaia_ns_summary} summarizes the Gaia NS inputs, and Section~\ref{subsec:sn_kicks} describes the instantaneous-SN calculation. We first fit Gaia NSs alone (Section \ref{subsec:gaia_only_method}).
We then fit Gaia NSs, Be+NS binaries, and isolated pulsars together using orbital constraints and birth rates for the binaries and velocity constraints from the pulsars to infer the kick distribution for all NS births (Section \ref{subsec:joint_method}).

\subsection{Gaia NS population}\label{subsec:gaia_ns_summary}

The Gaia NS population contains 27 detached NS--MS binary candidates identified in {\it Gaia} DR3 \citep{ElBadry2024b,ElBadry2024d,ElBadry2026}. The binaries span the following range of periods, eccentricities, and masses: $100<P_{\rm orb}/{\rm d}<1600$, $0<e<1$, $0.6<M_1/\msun<1.4$, and $1.25<M_{\rm NS}/\msun<2.2$.
Recently, \citet{Shariat2026GaiaNSDemographics} modeled the {\it Gaia} selection function and used a hierarchical Bayesian model \citep[see][]{Yamaguchi2026inprep} to constrain the intrinsic population demographics of Gaia NSs. They bias-correct the observed population using the {\it Gaia} selection function to infer the underlying distributions of orbital periods, eccentricities, neutron star masses, and companion masses.
The intrinsic period distribution is found to decrease mildly over the DR3-sensitive range, with power-law slope $\alpha_P=-0.75^{+0.37}_{-0.34}$.  
The intrinsic eccentricity distribution is modeled well by a beta distribution with mean $\mu_e=0.54^{+0.06}_{-0.05}$ and concentration $\kappa=4.8^{+2.7}_{-1.7}$. This distribution is
broad and weighted toward moderate eccentricities, with a median $e\approx0.55$ and middle-$50\%$ range $e\approx0.4$--$0.7$.
The companion masses peak near a solar mass, and the NS masses peak near $1.3~\msun$
similar to previous constraints on NS birth masses \citep[e.g.,][]{You2025}.

Another key result from \citet{Shariat2026GaiaNSDemographics} is the Gaia NS abundance. 
They infer a local space density of $\approx119~{\rm kpc^{-3}}$, which for observable lifetimes of $\sim1$--$14$~Gyr, corresponds to a Galactic birth rate of $\mathcal{R}_{\rm Gaia~NS}\sim10^{-6}$--$10^{-5}~{\rm yr^{-1}}$. 
Equivalently, only $1$ in $10^3-10^4$ NS births produce long-lived au-scale binaries with solar-type companions. 
Throughout this paper, we adopt a fiducial birth rate ratio $q_{\it Gaia}=\mathcal{R}_{\rm Gaia~NS}/\mathcal{R}_{\rm NS}=5\times10^{-4}$.
This term is used only as a weak rate prior to prevent pathological solutions that reproduce the Gaia NS orbits through an extremely fine-tuned, negligibly rare channel. The period--eccentricity distribution is still the main driver of the kick inference.

\subsubsection{Contamination from white dwarfs}\label{subsubsec:contamination}

Gaia NSs are NS {\it candidates} because the nature of the dark companion is inferred from its mass. 
Companions below the Chandrasekhar limit could be ultramassive WDs, and unresolved compact WD+WD binaries could also replicate a single NS-mass object \citep[e.g.,][]{ElBadry2024b,ElBadry2024d,ElBadry2026}.
Individual contaminants would not change the kick constraints for genuine Gaia NSs, but they would (a) reduce the true Gaia NS birth rate and (b) potentially bias the inferred kicks if they occupy a different part of the $P_{\rm orb}$--$e$ plane.  
The birth rate effect is likely subdominant, since the Gaia NS rate is already uncertain by a factor of $\sim2$ \citep{Shariat2026GaiaNSDemographics}. 
Single ultramassive-WD contaminants are also unlikely to dominate the kick inference because known {\it Gaia} WD+MS binaries tend to have relatively low eccentricities, unlike the Gaia NS candidates \citep{Shahaf2024, Yamaguchi2024,Yamaguchi2024b,Yamaguchi2025}. Unresolved compact WD+WD contaminants are harder to assess because their formation history and outer $P_{\rm orb}$--$e$ distribution remain poorly constrained \citep{Shariat2023, ElBadry2024b,ElBadry2024d}.

\citet{Shariat2026GaiaNSDemographics} repeated the population inference using only the $19$ Gaia NS candidates with $M_{\rm NS}>1.35~\msun$, where ultramassive-WD contamination is strongly reduced, and found similar period and eccentricity distributions\footnote{We also repeat the kick calculations for this subset, finding $1\sigma$ agreement.}.  For the remainder of this study, we follow the discovery papers in modeling the full sample as NS--MS candidates. Inferred kicks should likewise be interpreted as constraints on the genuine NS subset of the Gaia NS population.

\subsection{Supernova Kick Calculations}\label{subsec:sn_kicks}
We model the pre-supernova binary as {an exploding star of total mass $M_{\rm He}$} and a main-sequence companion of mass $M_2$ in a circular orbit\footnote{This is motivated by tidal circularization from the NS progenitor. Appendix \ref{app:presn_eccentricity} tests the effects of pre-supernova eccentricity.} with period $P_{\rm preSN}$. The {exploding star} instantaneously forms a $M_{\rm NS}=1.4~\msun$ NS and receives a natal kick $\mathbf{v}_{\rm kick}$, with mass lost $\Delta M=M_{\rm He}-M_{\rm NS}$.
The companion position and velocity are fixed during the explosion.  We then compute the new relative velocity, orbital energy, and angular momentum. 
A system is bound if the resulting orbital energy is negative and $0<e<1$.

{The au-scale orbits motivate stripped progenitors, since an extended red supergiant would likely interact with the companion \citep{ElBadry2024d}. However, we allow a broad progenitor mass range, $2.3<M_{\rm He}/\msun<12$, rather than fixing the mass loss to a small value. Appendix~\ref{app:high_mhe} explores weakly stripped progenitors, their lower survival efficiency, and the resulting required progenitor birth rate.}

The calculation follows the standard instantaneous-supernova formalism \citep{Blaauw1961,Hills1983,Kalogera1996,Kalogera1998, TaurisvandenHeuvel2023}.  The kick angle $\theta_v$ is measured between $\mathbf{v}_{\rm kick}$ and the orbital velocity of the exploding star $\mathbf{v}_{\rm He}$, which we sample isotropically.  Kicks with $90^\circ < \theta_v < 180^\circ$ oppose the pre-supernova motion and can reduce the post-supernova relative velocity, helping a wide binary remain bound with moderate eccentricity.  {The equations used to calculate the post-supernova orbits are given in Appendix~\ref{app:sn_formalism}.}

\subsection{Gaia NS fit}\label{subsec:gaia_only_method}

Our goal is to identify the pre-supernova period, {progenitor mass}, and kick speed distributions that reproduce the observed $P_{\rm orb}$--$e$ distribution of the $27$ Gaia NS candidates.  To make this comparison, we apply the supernova transformation, discard disrupted systems, and apply the {\it Gaia} astrometric selection function to the surviving binaries.

We calculate the Gaia DR3 astrometric selection function for NSs on a $23\times23$ grid in $\log_{10}P_{\rm final}$ and $e_{\rm final}$ over $100<P_{\rm final}/{\rm d}<1600$ (the relevant range for {\it Gaia} DR3) and $0<e_{\rm final}<1$, using $140{,}000$ \texttt{gaiamock} injection--recovery simulations from \citet{Shariat2026GaiaNSDemographics}.  
We smooth the weighted injection-recovery maps with a Gaussian kernel of size one grid cell and define the selection probability, $S_b$, in cell $b$.

For the Gaia NS-only model, we adopt a $1~\msun$ companion and model $\log_{10}(P_{\rm preSN}/{\rm d})$, $M_{\rm He}$, and $y=\ln(v_{\rm kick}/\kms)$
as independent normal distributions.  The period, {progenitor mass}, and kick speed draws are truncated to $1<P_{\rm preSN}/{\rm d}<3000$, $2.3<M_{\rm He}/\msun<12$, and $0.05<v_{\rm kick}/\kms<10^4$, respectively.  We therefore fit
\begin{equation}
    \Theta_{\rm Gaia}=(\mu_{\log P_{\rm preSN}},\sigma_{\log P_{\rm preSN}},
    \mu_{M_{\rm He}},\sigma_{M_{\rm He}},
    \mu_{\ln v},\sigma_{\ln v}).
\end{equation}

For each likelihood evaluation, we evolve $6\times10^4$ progenitors with the supernova equations and retain bound systems with $100<P_{\rm final}/{\rm d}<1600$ and $0<e_{\rm final}<1$.  We multiply the survivor histogram by $S_b$, smooth it with the same one-cell kernel, and normalize it to obtain the probability that a detected Gaia NS occupies cell $b$,
\begin{equation}
    p_{{\rm mod},{\it Gaia},b}(\Theta)
    =
    \frac{
    {\cal K}\left[S_b p_{{\rm int},{\it Gaia},b}(\Theta)\right]
    }{
    \sum_{b^\prime}
    {\cal K}\left[S_{b^\prime}p_{{\rm int},{\it Gaia},b^\prime}(\Theta)\right]
    },
    \label{eq:gaia_selected_model}
\end{equation}
where ${\cal K}$ denotes Gaussian smoothing and
$p_{{\rm int},{\it Gaia},b}(\Theta)$ is the intrinsic survivor distribution.

We use the fixed-count form of an inhomogeneous Poisson point-process likelihood conditioning on the observed number of Gaia NSs ($N=27$)\footnote{A Poisson likelihood would also contain the count-dependent terms $N\ln\Lambda-\Lambda$, where $\Lambda$ is the expected number of detected systems}:
\begin{equation}
\begin{aligned}
    \ln {\cal L}_{\it Gaia}
    &=
    \sum_b n_{{\rm obs},{\it Gaia},b}
    \ln p_{{\rm mod},{\it Gaia},b}(\Theta) \\
    &=
    \sum_{i=1}^{27}
    \ln p_{{\rm mod},{\it Gaia},b_i}(\Theta).
\end{aligned}
\label{eq:gaia_likelihood}
\end{equation}
where $n_{{\rm obs},{\it Gaia},b}$ is the number of Gaia NS candidates in
period--eccentricity cell $b$, and $b_i$ is the cell containing candidate $i$

The likelihood is larger when the Gaia-selected survivor distribution assigns high probability to the regions occupied by the observed systems.  Later in the joint fit, we add a separate rate term to constrain the population rate among all NSs (Section~\ref{subsec:joint_method}).
The posterior is sampled with \texttt{emcee} \citep{ForemanMackey2013} using $36$ walkers for $20{,}000$ steps, discarding the first $4{,}000$ steps as burn-in.

\subsection{Joint fit: Gaia NSs, Be+NSs, and isolated pulsars}\label{subsec:joint_method}

We fit the $27$ Gaia NS candidates and $23$ Be+NS binaries with one shared two-component kick velocity distribution, while allowing their pre-supernova period and {progenitor mass} distributions to differ.
The binary orbits and birth rates constrain the shape and weight of the low-kick component, and velocities of young isolated pulsars constrain the high-kick component \citep{Disberg2025}.

\subsubsection{The Be+NS Population}\label{subsubsec:be_ns_pop}
High-mass Be X-ray binaries contain an NS orbiting a classical Be star: a rapidly rotating B star with a viscous decretion disk \citep{Reig2011}. During its orbit, the NS accretes from the Be disk, often near periastron.
The X-ray pulsations trace the NS spin period, and measuring their arrival-time delays yields precise orbital periods and eccentricities. 
Wide Be X-ray binaries ($P_{\rm orb} \gtrsim 5$~days) experience negligible tides over the Be-star lifetime \citep[e.g.,][]{Bashi2023}. Thus, like Gaia NSs, they are clean probes of NS natal kicks because their present-day orbits have remained virtually unchanged since the explosion.  
Their pre-supernova evolution is also comparatively
well understood: mass transfer stripped the NS progenitor before core collapse and spun up the present Be star \citep{Reig2011,Tauris2017,Valli2025}. 

For fitting the kicks, we follow \citet{Valli2025} in using a Galactic Be-XRB sample with 23 systems.  Their sample requires a confirmed NS, a classical Be companion, and accurately measured $P_{\rm orb}$ and $e$.  The systems are restricted to those with
\begin{equation}\label{eq:peri_range}
    30<a_{\rm post}(1-e_{\rm post})/{\rm \rsun}<450.
\end{equation}
The lower limit is wide enough for a Be-star disk to form and for tidal circularization to be minimal on stellar evolution timescales, and the upper limit keeps the NS close enough to accrete from the disk and appear as a Be X-ray binary. Note that while the selection biases of Be X-ray binaries are poorly understood relative to the Gaia NS population, \citet{Valli2025} identify no major biases in the selected  $P_{\rm orb}$ and $e$ range relevant for the kick analysis.

{The Be+NS sample can be visually separated into a low-eccentricity group and a high-eccentricity sequence with an $e=0.25$ dividing line. \citet{Valli2025} modeled the kicks of these branches separately. We instead take the agnostic approach in fitting the sample altogether. The two results are compared in Section~\ref{subsubsec:comp_to_BeNS}.}

\subsubsection{The isolated-pulsar population}\label{subsubsec:dm25_pulsars}

Young isolated pulsars provide the high-velocity component for our global kick model.  \citet{Disberg2025} modeled the 3D velocities of young isolated pulsars relative to their local standard of rest using proper motions and distance measurements.  Their fiducial sample contains $48$ pulsars with parallax distances and characteristic ages $\tau_c\leq10$ Myr, young enough that their present-day velocities should remain close to their birth velocities.  They found that this sample is consistent with
\begin{equation}
    \ln(v_{\rm kick}/\kms)\sim
    {\cal N}(\mu_{\ln v,{\rm pulsar}},\sigma_{\ln v,{\rm pulsar}}),
    \label{eq:dm25_kick_law}
\end{equation}
with $\mu_{\ln v,{\rm pulsar}}=5.60\pm0.12$ and $\sigma_{\ln v,{\rm pulsar}}=0.68\pm0.10$ \citep{Disberg2025}, corresponding to a mode near $170~\kms$.
For the fits, we propagate these reported uncertainties by treating both parameters as nuisance parameters with independent Gaussian priors,
\begin{equation}
\begin{split}
\mu_{\ln v,{\rm pulsar}} & \sim {\cal N}(5.60,0.12^2),\\
\sigma_{\ln v,{\rm pulsar}} &\sim {\cal N}(0.68,0.10^2),
\qquad \sigma_{\ln v,{\rm pulsar}}>0.
\end{split}
\label{eq:dm25_priors}
\end{equation}

\subsubsection{Joint fit}\label{subsubsec:joint_fit}

For both binary populations, we assume circular pre-supernova orbits\footnote{While prior binary interactions can circularize the orbit, perturbations from a tertiary star may maintain or excite eccentricity \citep[e.g.,][]{Rose2019,Kummer2023,Shariat2025b, Sciarini2025, Bruenech2025, Stegmann2025}. Appendix \ref{app:presn_eccentricity} tests the effect of initial eccentricity.
} and isotropic kick directions. 
The Gaia NS calculation assumes a $1~\msun$ companion and $1.4~\msun$ NS, as in Section \ref{subsec:gaia_only_method}.  
{For Be+NS binaries, we adopt a $15~\msun$ companion and retain bound systems satisfying Equation~\ref{eq:peri_range}, following \citet{Valli2025}.}
{We allow the pre-supernova periods and progenitor masses to differ between the Gaia NS and Be+NS populations, since their formation histories are likely different (Section~\ref{subsec:physical_interpretation}).} 
{The actual natal kick is thought to arise from asymmetries during NS formation \citep{Janka2017,JankaKresse2024}. We assume a shared parent kick velocity distribution for Gaia NSs, Be+NS binaries, and young isolated pulsars.}

{The universal kick velocity law is assumed to be a two-component model in $y=\ln(v_{\rm kick}/\kms)$:}
\begin{equation}
\begin{split}
    p(y)=f_{\rm low}\,p_{\rm low}(y)
 +(1-f_{\rm low})\,p_{\rm pulsar}(y),
\end{split}
\label{eq:joint_kick_law}
\end{equation}
truncated to $0.05<v_{\rm kick}/\kms<1000$. We assume the low-kick branch follows a lognormal distribution, 
\begin{equation}
p_{\rm low}(y)={\cal N}(y\mid\mu_{\ln v,{\rm low}},\sigma_{\ln v,{\rm low}}),
\end{equation}
and $p_{\rm pulsar}(y) \propto{\cal N}(y\mid\mu_{\ln v,{\rm pulsar}},\sigma_{\ln v,{\rm pulsar}})$. Each component is normalized separately over the truncated interval, and the velocity density is $p(v)=p(y)/v$. 
{We use lognormal distributions as a simple model for positive kick speeds with a free median and width, following \citet{Disberg2025}.}

The fit has 13 free parameters,
\begin{equation}
\begin{aligned}
\Theta=(&\mu_{\log P,G},\sigma_{\log P,G},\mu_{M,G},\sigma_{M,G},\\
&\mu_{\log P,B},\sigma_{\log P,B},\mu_{M,B},\sigma_{M,B},\\
&\mu_{\ln v,{\rm low}},\sigma_{\ln v,{\rm low}},f_{\rm low},\\
&\mu_{\ln v,{\rm pulsar}},\sigma_{\ln v,{\rm pulsar}}).
\end{aligned}
\end{equation}
We use uniform priors over the range: $1<P_{\rm preSN}/{\rm d}<3000$, $2.3<M_{\rm He}/\msun<12$, $0<\mu_{\ln v,{\rm low}}<4.8$, $0.05<\sigma_{\ln v,{\rm low}}<1.8$, and $0<f_{\rm low}<1$. The priors for $\mu_{\ln v,{\rm pulsar}}$ and $\sigma_{\ln v,{\rm pulsar}}$ are provided by Equation \eqref{eq:dm25_priors}.

For Gaia NSs, we use the likelihood from Equation~\ref{eq:gaia_likelihood}, sampling $3\times10^4$ progenitors per evaluation.
For Be+NS binaries, we use an analogous likelihood. After sampling $3\times10^4$ progenitors, we retain bound systems satisfying Equation~\ref{eq:peri_range} and compare their normalized, smoothed $P_{\rm orb}$--$e$ distribution to the $23$ observed systems:
\begin{equation}
\begin{aligned}
    \ln {\cal L}_{\rm Be}
    &=
    \sum_b n_{{\rm obs},{\rm Be},b}
    \ln p_{{\rm mod},{\rm Be},b}(\Theta) \\
    &=
    \sum_{j=1}^{23}
    \ln p_{{\rm mod},{\rm Be},b_j}(\Theta),
\end{aligned}
\label{eq:be_likelihood}
\end{equation}
Here $n_{{\rm obs},{\rm Be},b}$ is the number of observed systems in cell $b$, and $p_{{\rm mod},{\rm Be},b}$ is the normalized model probability in that cell.  Following \citet{Valli2025}, we assume no additional selection function for the Be+NS sample.  

Lastly, we introduce a term constraining the birth rates of each population.  For each population $c$, the kick simulation gives a survivor efficiency $\eta_c(\Theta)$: the fraction of trial pre-supernova binaries that remain bound and pass the Gaia NS or Be+NS selection.  
For population $c$, let $q_c$ be its inferred birth rate relative to the total NS birth rate. If a model produces that population with survivor efficiency $\eta_c(\Theta)$, then the required fraction of all NS births that pass through the corresponding progenitor channel is
\begin{equation}
    f_{{\rm ch},c}(\Theta)=\frac{q_c}{\eta_c(\Theta)}.
    \label{eq:channel_fraction}
\end{equation}
We adopt $q_{\it Gaia}=4.74\times10^{-4}$ and $q_{\rm Be}=3\times10^{-3}$ in the likelihood \citep{Shariat2026GaiaNSDemographics}, both uncertain within a factor of two.
The Galactic Be-XRB sample is affected by distance, extinction, duty cycle, and X-ray selection biases \citep{Fortin2023}, so the estimate is anchored to the less-biased LMC Be-XRB population \citep{Antoniou2016}.  

We represent the uncertainty in the rates with a Gaussian constraint on $\log_{10}f_{\rm ch}$,
\begin{equation}
\begin{split}
    \ln {\cal L}_{\rm rate}={}&-\frac{1}{2}
    \sum_{c\in\{{\it Gaia},{\rm Be}\}}
    \left[\frac{\Delta_c}{\sigma_{\rm rate}}\right]^2 
    ,\\
    \Delta_c={}&\log_{10}(f_{{\rm ch},c})
    -\log_{10}(f_{{\rm ch},c,0}),
\end{split}
\label{eq:kick_likelihood_rate}
\end{equation}
with $f_{{\rm ch},{\it Gaia},0}=10^{-2}$, $f_{{\rm ch},{\rm Be},0}=3\times10^{-2}$, and $\sigma_{\rm rate}=0.2$ dex. This likelihood term constrains the channel fraction implied by the kick model ($f_{{\rm ch},c}$) around the adopted expectation ($f_{{\rm ch},c,0}$).
{This rate likelihood term disfavors solutions where the required Gaia NS or Be+NS channel would represent an implausibly small or implausibly large fraction of all NS births. }

The total likelihood is
\begin{equation}
    \ln {\cal L}_{\rm tot}
    =
    \ln {\cal L}_{\it Gaia}
    +\ln {\cal L}_{\rm Be}
    +\ln {\cal L}_{\rm rate}.
    \label{eq:joint_likelihood}
\end{equation}

We sample this posterior with \texttt{emcee} \citep{ForemanMackey2013}, using $48$ walkers for $6000$ steps and discarding the first $1800$ steps.

\section{Natal Kick Constraints}\label{sec:kick_constraints}

\subsection{Gaia NS-only kicks}\label{sec:gaia_only}

The Gaia NS fit provides the {\it parent} distribution of the pre-supernova periods, {progenitor masses}, and kick speeds assigned to He-star + MS progenitors before the supernova.  It also provides the {\it survivor-weighted} distribution: the subset of the kick prescription that produces surviving Gaia NS binaries (see Appendix \ref{app:survivor_distributions}).

The inferred {\it parent} kick law is a lognormal distribution with
\begin{equation}
    \mu_{\ln v}=3.19^{+0.85}_{-1.34},
    \qquad
    \sigma_{\ln v}=0.70^{+0.92}_{-0.52},
\end{equation}
corresponding to a parent speed
$v_{\rm kick}=24.2^{+32.6}_{-17.9}~\kms$, below the characteristic speeds inferred for young isolated pulsars \citep[e.g.,][]{Hobbs2005}.  The Gaia NS orbits alone allow for broad progenitor distributions.  The pre-SN period distribution favors most systems having $P_{\rm preSN}\sim100$--$1{,}000$~d before the explosion.
The {progenitor masses} are similarly broad and allow progenitor masses from $\sim2.3$--$8~\msun$ with a peak near $\sim5~\msun$. Table~\ref{tab:gaia_only_fit_params} summarizes the fitted parameters.

The survivor-weighted distributions (Appendix \ref{app:survivor_distributions}) are shifted further toward favorable kicks. 
Binaries that survive as Gaia NSs select close au-scale initial orbits ($P_{\rm preSN}=220^{+110}_{-70}$~d), modest mass loss ($M_{\rm He}=4.4^{+2.0}_{-1.3}~\msun$), and relatively low kick speeds ($v_{\rm kick}=29^{+14}_{-20}~\kms$). 
The preferred kick directions oppose the pre-supernova orbital motion ($\theta_v=143^{+8}_{-16}$~deg), reducing the post-supernova relative velocity and allowing an au-scale binary to remain bound with moderate eccentricity \citep[see also][]{ElBadry2024d}. 

\vspace{-0.5cm}
\begin{deluxetable}{lc}
\tablecaption{Parent distribution parameters for the Gaia-only model.\label{tab:gaia_only_fit_params}}
\tablehead{
\colhead{Parameter} & \colhead{Posterior median}
}
\startdata
$\mu_{\log_{10}P_{\rm preSN}}$ & $2.42^{+0.48}_{-0.28}$ \\
$\sigma_{\log_{10}P_{\rm preSN}}$ & $0.51^{+0.37}_{-0.19}$ \\
$\mu_{M_{\rm He}}/\msun$ & $5.05^{+1.70}_{-2.03}$ \\
$\sigma_{M_{\rm He}}/\msun$ & $0.61^{+1.19}_{-0.44}$ \\
$\mu_{\ln v}$ & $3.19^{+0.85}_{-1.34}$ \\
$\sigma_{\ln v}$ & $0.70^{+0.92}_{-0.52}$ \\
\enddata
\end{deluxetable}

Instantaneous symmetric mass loss produces 
\begin{equation}\label{eq:blaauw_ecc}
    e_{\rm Blaauw}=\frac{\Delta M}{M_{\rm post}},
    \qquad
    \Delta M < M_{\rm post}\quad{\rm (bound)},
\end{equation}
where $M_{\rm post}=M_{\rm NS}+M_2$.  The binary becomes unbound when $e_{\rm Blaauw}\geq1$, or equivalently when $\Delta M\geq M_{\rm post}$. The latter is the classical result that an initially circular binary must retain more than half of its pre-supernova mass to remain bound \citep[e.g.,][]{Blaauw1961,Hills1983}. 

In the Gaia NS calculation, $M_{\rm post}=2.4~\msun$. Even near the minimum allowed mass loss, $\Delta M\approx1~\msun$, gives $e_{\rm Blaauw}\approx0.4$ without any extra natal kick, while the best-fit median $M_{\rm He}\approx5.1~\msun$ would unbind the binary.
The low- and moderate-eccentricity Gaia NSs require a nonzero natal kick that partly cancels the Blaauw impulse and keeps the binary bound. Compared to isolated-pulsar velocities, these required kicks are small.

\begin{figure*}
    \centering
    \includegraphics[width=0.92\linewidth]{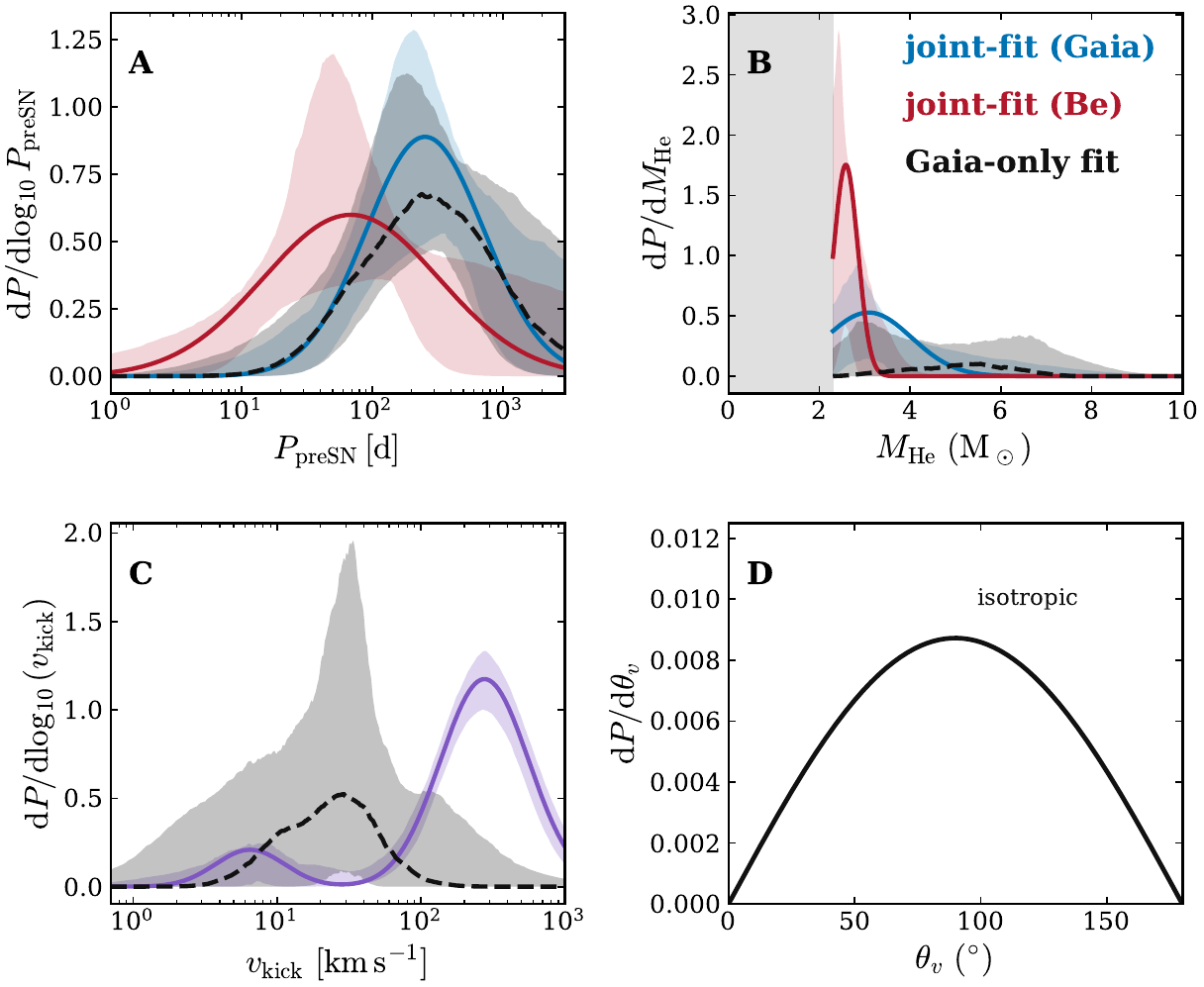}
    \caption{Parent kick distributions for Gaia NSs and Be+NS binaries. Panels A--D show pre-supernova orbital periods ($P_{\rm preSN}$), progenitor masses ($M_{\rm He}$), kick velocities ($v_{\rm kick}$), and kick directions ($\theta_v$).
    Dashed black curves ($1\sigma$ shaded gray) show the resulting kick law from fitting Gaia NS orbits only. 
    Blue and red curves show the results for Gaia NS and Be+NS binaries after fitting both populations simultaneously with isolated pulsars.
    The survivor-weighted distributions are shown in Appendix \ref{app:survivor_distributions}. The shared, bimodal $v_{\rm kick}$ distribution can simultaneously produce all three populations, although Gaia NSs and Be+NSs generally arise from different progenitors and kicks. }\label{fig:parent_fourpanel}
\end{figure*}

\subsection{A concordance model of NS natal kicks}\label{sec:joint}

\vspace{-0.5cm}
\begin{deluxetable}{lc}
\tablecaption{Parent distribution parameters for the unified kick model.\label{tab:joint_fit_params}}
\tablehead{
\colhead{Parameter} & \colhead{Best-fit}
}
\startdata
\sidehead{{\it Gaia NS progenitors}}
$\mu_{\log_{10}P_{\rm preSN}}$ & $2.41^{+0.42}_{-0.16}$ \\
$\sigma_{\log_{10}P_{\rm preSN}}$ & $0.45^{+0.38}_{-0.16}$ \\
$\mu_{M_{\rm He}}/\msun$ & $3.08^{+1.34}_{-0.48}$ \\
$\sigma_{M_{\rm He}}/\msun$ & $0.95^{+1.26}_{-0.64}$ \\
\sidehead{{\it Be+NS progenitors}}
$\mu_{\log_{10}P_{\rm preSN}}$ & $1.84^{+0.70}_{-0.22}$ \\
$\sigma_{\log_{10}P_{\rm preSN}}$ & $0.67^{+0.66}_{-0.35}$ \\
$\mu_{M_{\rm He}}/\msun$ & $2.58^{+0.35}_{-0.21}$ \\
$\sigma_{M_{\rm He}}/\msun$ & $0.27^{+0.31}_{-0.14}$ \\
\sidehead{{\it Shared low-kick component}}
$f_{\rm low}$ & $0.126^{+0.082}_{-0.044}$ \\
$\mu_{\ln v,{\rm low}}$ & $1.87^{+0.65}_{-0.53}$ \\
$\sigma_{\ln v,{\rm low}}$ & $0.55^{+0.85}_{-0.44}$ \\
\sidehead{{\it Pulsar-constrained high-kick component}}
$\mu_{\ln v,{\rm pulsar}}$ & $5.62^{+0.12}_{-0.12}$ \\
$\sigma_{\ln v,{\rm pulsar}}$ & $0.71^{+0.09}_{-0.09}$ \\
\enddata
\end{deluxetable}

Figure~\ref{fig:parent_fourpanel} shows the parent distributions from the joint fit, designed to be consistent with the Gaia NS, Be+NS, and isolated pulsar populations  (results summarized in Table~\ref{tab:joint_fit_params}).  The corresponding survivor-weighted distributions (i.e., after selecting only the surviving binaries) appear in Appendix~\ref{app:survivor_distributions}.

The inferred low-kick component has 
\begin{equation}
    \mu_{\ln v,{\rm low}}=1.87^{+0.65}_{-0.53}
    \qquad
    \sigma_{\ln v,{\rm low}}=0.55^{+0.85}_{-0.44},
\end{equation}
corresponding to a median kick speed of $6.5^{+5.9}_{-2.7}~\kms$. 
The fraction of all NS births that receive low kicks is $f_{\rm low}=0.126^{+0.082}_{-0.044}$, implying that roughly $1$ in $10$ NSs form with low kicks. 
A short Python function for sampling the global natal kick
distribution at our inferred parameter values is available
online.\footnote{\url{https://github.com/cheyanneshariat/ns-kick-sampler}}

The survivor-weighted distributions are more tightly constrained than the individual parent-component parameters.
Because low kicks preferentially preserve wide binaries, $87\%$ ($74$--$94\%$) of selected Gaia NSs arise from the low component, compared with $46\%$ ($35$--$62\%$) of selected Be+NS binaries. The median kick speeds of the selected binary populations are $6.8^{+12.0}_{-2.6}~\kms$ and $47.9^{+27.5}_{-29.4}~\kms$, respectively.
Figure~\ref{fig:survivor_fourpanel_appendix} shows how survival and the sample selection transform the parent law into these observed binary populations.

\begin{figure*}
    \centering
    \includegraphics[width=0.94\linewidth]{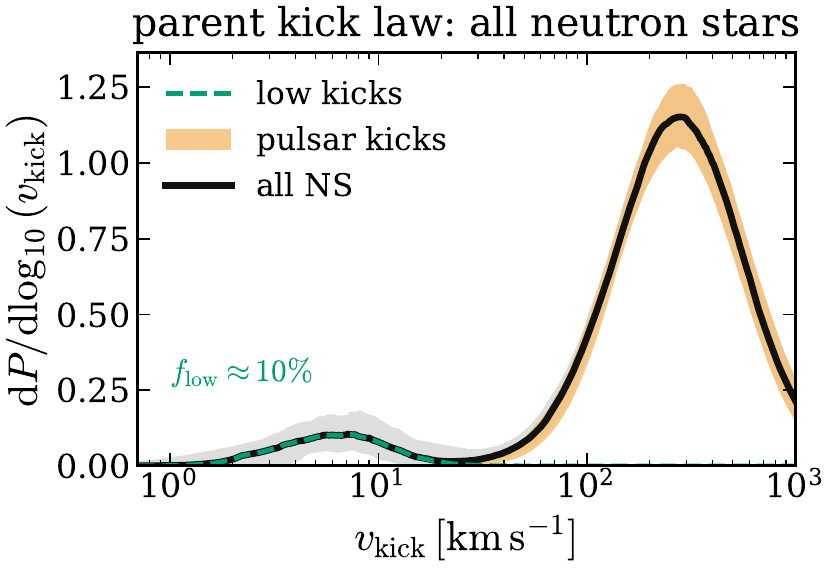}
    \caption{\textbf{Inferred parent natal kick distribution for all NSs.}  The black curve shows the unified kick law with $1\sigma$ range shaded gray. The orange region shows the high-kick component, whose parameters are informed by the velocities of young isolated pulsars \citep{Disberg2025}.  The green dashed curve shows the low-kick contribution inferred from Gaia NS and Be+NS binaries.  The black distribution integrates to unity, while the area under the green curve is $f_{\rm low}\approx10\%$. 
    Low kicks dominate the surviving binary populations, but remain a minority of all NS birth kicks.}
    \label{fig:parent_kick_law}
\end{figure*}

Figure~\ref{fig:parent_kick_law} shows the resulting NS kick distribution for all NS births.  
The full parent distribution is dominated by the isolated-pulsar law ($v_{\rm kick}\gtrsim100~\kms$), with an additional $\sim10\%$ component at lower velocities ($v_{\rm kick}\lesssim50~\kms$).

Our analysis does not constrain the physical origin of this component, although possible channels include electron-capture, ultra-stripped, and weak iron-core-collapse supernovae (see Section \ref{subsec:physical_interpretation} for a discussion).

\subsubsection{Survival efficiencies and birth rates}\label{subsubsec:survival_rates}

The model predicts that $4.0^{+1.9}_{-1.6}\%$ of Gaia NS progenitor binaries and $21.4^{+7.7}_{-6.5}\%$ of Be+NS progenitor binaries remain bound after the supernova and pass the corresponding sample selection. The Be+NS efficiency is comparable to the $\sim14\%$--$25\%$ survival fractions predicted for massive binaries \citep[e.g.,][]{Renzo2019,Kochanek2019}, 
Combined with the observed birth rates (Equation \ref{eq:channel_fraction}), these efficiencies imply that $f_{{\rm ch},{\it Gaia}}=1.2^{+1.3}_{-0.6}\%$ of all NS births pass through the modeled Gaia NS progenitor phase and $f_{{\rm ch},{\rm Be}}=1.4^{+0.6}_{-0.4}\%$ pass through the modeled Be+NS progenitor phase. 
These progenitor channels refer to the pre-SN binary configurations: $\sim1$~au He+MS binaries for Gaia NSs and $\sim0.1$~au He+Be binaries for Be+NSs (e.g., Figure~\ref{fig:parent_fourpanel}).

\subsubsection{Progenitor binaries}\label{subsubsec:joint_progenitors}

The Gaia NS progenitor mass and $f_{\rm low}$ are correlated. 
Larger progenitor masses produce larger Blaauw kicks, requiring higher natal kick velocities to partly cancel them. These higher natal kicks lead to more unbound systems, and thus require a larger $f_{\rm low}$ to replenish the supply of Gaia NS progenitors.
This can be shown from the fits, where conditioning the broad posterior on $\mu_{M_{\rm He,G}}=2.3$--$3$, $3$--$4$, $4$--$5$, and $5$--$8~\msun$ produces $f_{\rm low}=11.0^{+5.1}_{-3.9}\%$, $14.0^{+6.4}_{-4.8}\%$, $15.5^{+15.1}_{-6.5}\%$, and $18.1^{+18.6}_{-8.9}\%$, respectively.

The preference for low-$M_{\rm He}$ in the Be+NS population is mainly driven by the low-eccentricity Be+NS binaries and the assumption that the Be+NS {progenitor mass} distribution is single-peaked.  For a $15~\msun$ Be companion, a strongly stripped helium star with $M_{\rm He}\approx2.6~\msun$ loses only $\Delta M\approx1.2~\msun$, giving a Blaauw eccentricity of $\Delta M/M_{\rm post}\approx0.07$ with $v_{\rm kick}=0$ (Equation \ref{eq:blaauw_ecc}). Thus, very small kicks naturally populate the low-eccentricity Be+NS branch, similar to the low-kick, low-mass-loss branch found by \citet{Valli2025}.  
With the same low-$M_{\rm He}$, the eccentric Be+NS branch can be produced by the high-kick component, $v_{\rm kick}\gtrsim50~\kms$.
Thus, one peak in the $M_{\rm He}$ distribution at $M_{\rm He} \approx 2.5~\msun$ can account for both Be+NS branches\footnote{We also test a more flexible model that allows for two peaks in the $M_{\rm He}$ distribution, finding that the maximum likelihood changes by $<0.5$, while the additional parameters increase the AIC by $\Delta{\rm AIC}=12.9$, favoring the adopted single-component model.}. 

In the joint fit, surviving Gaia NS binaries come from au-scale pre-supernova binaries, typically $0.1$--$1$~au. This predicts a population of intermediate-mass (possibly stripped) stars ($M_{\rm He}=2.3$--$8~\msun$) with $\sim1~\msun$ main-sequence companions with $P_{\rm orb}\approx50$--$1000$~d (Figure \ref{fig:parent_fourpanel}). 

The Be+NS binaries have a tighter $P_{\rm preSN}$ distribution, with median $P_{\rm preSN}\approx45$ d and a typical range of $\approx20$--$200$ d. These are consistent with previous kick constraints and with the standard formation picture of Be XRBs \citep{Packet1981,Pols1991,Reig2011,Vinciguerra2020,Liu2024,Valli2025, LopezOller2026}.

\subsubsection{Posterior predictive checks}\label{subsec:pe_check}

For each posterior draw, we generate mock Gaia NS and Be+NS populations, keep surviving binaries, apply observational selection effects, and map the resulting systems onto the $P_{\rm orb}$--$e$ plane.  Repeating this calculation across the posterior gives posterior-predictive distributions.

\begin{figure}
    \centering
    \includegraphics[width=0.99\linewidth]{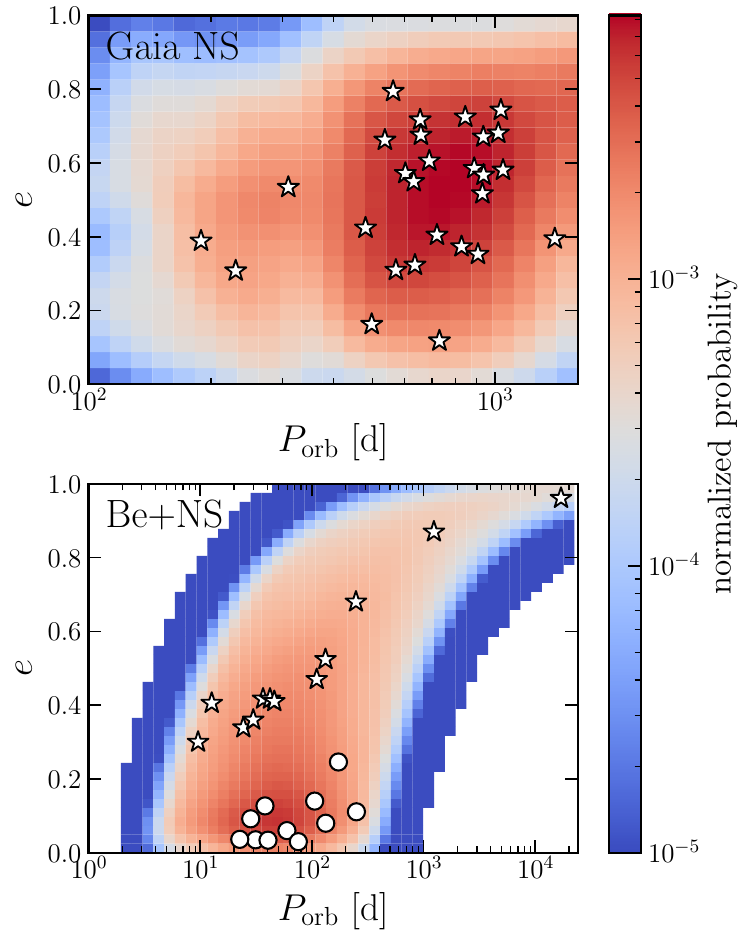}
    \caption{Posterior predictive periods and eccentricities from the joint Gaia--Be kick model.  The top panel shows Gaia NS survivors over the period range, and the bottom panel shows Be+NS survivors over the Be-XRB period range. The top panel includes the {\it Gaia} astrometric selection function, while the bottom assumes no additional survey-selection effects. 
    Each panel is normalized to unit probability.  White stars mark the observed Gaia systems; open circles and stars mark the low- and high-eccentricity Be+NS systems (fit together). Applying the parent kick law (Figure~\ref{fig:parent_fourpanel}) reproduces the broad range of Gaia NS orbits and both Be+NS branches reasonably well.}
    \label{fig:joint_pe_check}
\end{figure}

Figure~\ref{fig:joint_pe_check} shows the posterior-predicted $P_{\rm orb}$--$e$ distribution under the shared kick law.
{For Gaia NSs, we account for each simulated binary's probability of detection by Gaia DR3, then draw mock catalogs containing $N=27$ detected systems.}
These catalogs have median period $731$~d ($648$--$824$~d) and median eccentricity $0.54$ ($0.46$--$0.59$), compared to $690$~d and $0.549$ for the observed sample (top panel). The model predicts $4$ ($2$--$6$) systems with $e<0.3$, consistent with the $2$ observed.
{It also predicts few Gaia detections at short periods and low eccentricities, consistent with the absence of observed systems in this region.}
\footnote{{Tests with reduced completeness at high eccentricity give consistent low-kick constraints.}}
The $\sim10\times$ larger DR4 sample predicted by \citet{Shariat2026GaiaNSDemographics} will test whether this region remains sparsely populated.

For Be+NS binaries, mock $23$-system catalogs have median period $77$~d ($49$--$120$~d)  and median eccentricity $0.27$ ($0.17$--$0.38$), compared to the observed values of $46$~d and $0.30$.  The model predicts $9$ ($6$--$12$) systems with $e<0.2$, consistent with the $10$ observed. The high-eccentricity tail is also consistent, predicting $4$ ($2$--$6$) systems with $e>0.6$ compared to $3$ observed.  

{Isotropic kick directions with circular progenitor orbits reproduce the sequence reasonably well, simultaneously with the low-$e$ binaries, at the current level of statistical power (Section \ref{subsubsec:comp_to_BeNS} compares alternative interpretations).
Our model also predicts Be+NS binaries in the upper region of the $P$--$e$ plane, where none of the Be+NS binaries in the \citet{Valli2025} sample reside, although such binaries do exist (discussed in Section \ref{subapp:toomre_bens}).}

\subsubsection{A unimodal kick distribution?}\label{subsec:broad_unimodal_main}

Young pulsar velocities have been modeled with a double-peaked kick velocity distribution \citep[e.g.,][]{FaucherGiguere2006,Verbunt2017,Igoshev2020}, but \citet{Disberg2025} recently argued that a unimodal lognormal distribution is statistically consistent with the current young-pulsar sample. Gaia NSs and Be+NS binaries are rare compared to isolated NSs \citep[e.g.,][]{Shariat2026GaiaNSDemographics}, raising the question of whether they are simply drawn from the low-velocity tail of the young pulsar distribution.

{We test this hypothesis by assuming a single lognormal kick distribution,
\begin{equation}
    \ln(v_{\rm kick}/\kms)\sim \mathcal{N}(\mu_{\ln v},\sigma_{\ln v}),
\end{equation}
and test three variations: the alternative $\tau_c\leq40$ Myr fit from \citet{Disberg2025}, $(\mu_{\ln v},\sigma_{\ln v})=(5.67,0.59)$; their fiducial $\tau_c\leq10$ Myr fit, $(5.60,0.68)$; and a broadened low-velocity stress test, $(5.48,0.88)$. All three are consistent within $2\sigma$ with the pulsar space velocities \citep{Disberg2025}.

We adopt circular He+MS progenitors with $M_2=1~\msun$ and $M_{\rm NS}=1.4~\msun$ and sample pre-supernova periods and progenitor masses from the best-fit model in Section \ref{sec:gaia_only}. 
For each kick law, we draw $2.4\times10^5$ explosions with isotropic kick directions, apply the supernova transformation, and retain bound systems with finite eccentricity and $100<P_{\rm orb}/{\rm d}<1600$. We define the Gaia NS-like efficiency as $\eta=N_{\rm bound,\,Gaia}/N_{\rm MC}$ and apply the DR3 selection weights $S_{ij}=N_{\rm selected,ij}/N_{\rm injected,ij}$ from the injection--recovery tests of \citet{Shariat2026GaiaNSDemographics}.

Single-lognormal models produce some Gaia NSs, but place too many of them at short periods and high eccentricities (Figure~\ref{fig:single_lognormal_appendix}). In contrast, no observed Gaia NSs reside above the cyan line,
\begin{equation}
    e > 0.20 + 0.89\left[\log_{10}(P_{\rm orb}/{\rm d})-2.00\right].
\end{equation}
We quantify this difference by computing, after applying the DR3 selection function, the fraction of selected survivors above this line.

\begin{figure*}
    \centering
    \includegraphics[width=0.99\linewidth]{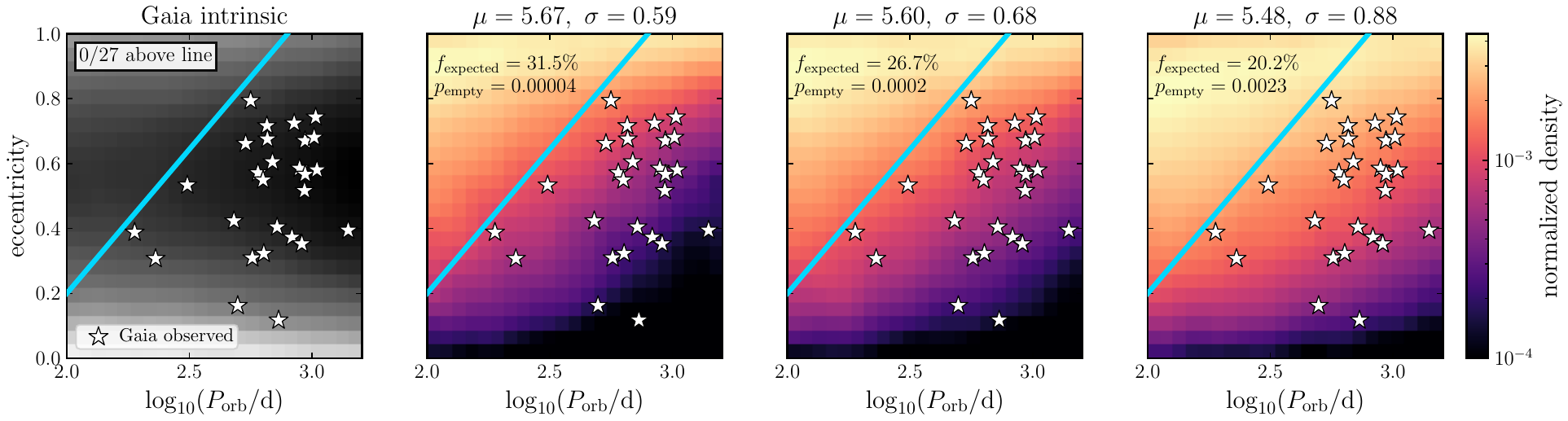}
    \caption{Predicted $P$--$e$ distributions for single-lognormal kick laws. The left panel shows the intrinsic Gaia NS distribution \citep{Shariat2026GaiaNSDemographics} with the 27 observed candidates (white stars). The remaining panels show predictions for the alternative $\tau_c\leq40$ Myr and fiducial $\tau_c\leq10$ Myr fits of \citet{Disberg2025}, and a broadened low-velocity distribution. The adopted lognormal parameters are shown above each panel. The cyan line bounds a region containing no observed Gaia NSs. Here, $f_{\rm expected}$ is the predicted fraction of detected systems above the line, and $p_{\rm empty}$ is the probability of observing zero systems there in a sample of 27. Even the broadened distribution places too many selected survivors in this region.}
    \label{fig:single_lognormal_appendix}
\end{figure*}

The three kick laws predict that $20$--$32\%$ of {\it detected} Gaia NSs should lie above this line, while $0/27$ do. The fiducial \citet{Disberg2025} pulsar lognormal predicts $f_{\rm empty}=0.37$, so the probability of drawing zero systems in this region in Gaia DR3 is
\begin{equation}
    p_{\rm empty}=(1-f_{\rm empty})^{27}\approx2.3\times10^{-5}.
\end{equation}
A deliberately broadened pulsar-like distribution still gives $f_{\rm empty}=0.20$ and $p_{\rm empty}=2.3\times10^{-3}$. The problem is therefore not only the low survivor fraction, but that the surviving binaries occupy the wrong part of the $P$--$e$ plane. Under these progenitor assumptions, Gaia NSs favor a distinct low-kick component, beyond the pulsar contribution. Related constraints from Be+NS and double NS binaries are discussed in Section~\ref{subsec:previous_kick_constraints} \citep{Valli2025,Grichener2026,DisbergMandelHirai2026}.}

\section{Kinematics}\label{sec:kinematics}
{An impulsive kick changes both the binary's orbit and its center-of-mass velocity, altering its kinematic motion through the Galaxy.
The present-day space velocities of post-kick binaries thus provide a complementary constraint on natal kicks.
We use observed kinematics of Gaia NSs (Section \ref{subsec:toomre_main}), NS-LMXBs (Section \ref{subsec:lmxb_compare}), and Be+NS binaries (Section \ref{subsec:bens_kinematics}) to independently test the unified NS kick model.}

\subsection{Gaia NS kinematics}\label{subsec:toomre_main}
\subsubsection{Comparison to nearby stars}

We compare the 3D velocities of Gaia NS candidates to their local stellar population following the approach of \citet{Nagarajan2025a} for Gaia black holes.
{Appendix~\ref{app:toomre} details the reference sample.}
For each Gaia NS candidate, we compute Galactocentric cylindrical velocities and compare them to nearby {\it Gaia} DR3 stars.  Comparing to local stars minimizes the effects of Galactic rotation and kinematic heating, which can be substantial for old NS populations such as Gaia NSs. 
Any significant excess in spatial velocity with respect to the local population can be interpreted as a lower limit to the velocity kick received \citep[e.g.,][]{Nagarajan2025a}: 
\begin{equation}
    v_{\rm kick,min} = {\rm max}(0,V - V_{68\%}),
\end{equation}
where the $68\%$ threshold ($V_{68\%}$) is chosen as a $1\sigma$ estimate following \citet{Nagarajan2025a}. A more conservative choice would be to use $V_{95\%}$, but we find that qualitative results are not highly sensitive to this choice. The values of $v_{\rm kick,min}$, $V_{68\%}$, and $V_{95\%}$ for each candidate are listed in Table \ref{tab:gaia_ns_kinematics}. 
Three Gaia NSs (J1739+4502, J0152$-$2049, and J1432$-$1021) reside on halo-like orbits and have low metallicities (${\rm [Fe/H]}<-1.2$), suggesting that their high space velocities mainly reflect halo membership \citep{ElBadry2024b}. We report them in Table~\ref{tab:gaia_ns_kinematics} but exclude them from the 24-system disk comparison and posterior-predictive calculation.

\begin{figure*}
    \centering
    \includegraphics[width=0.95\linewidth]{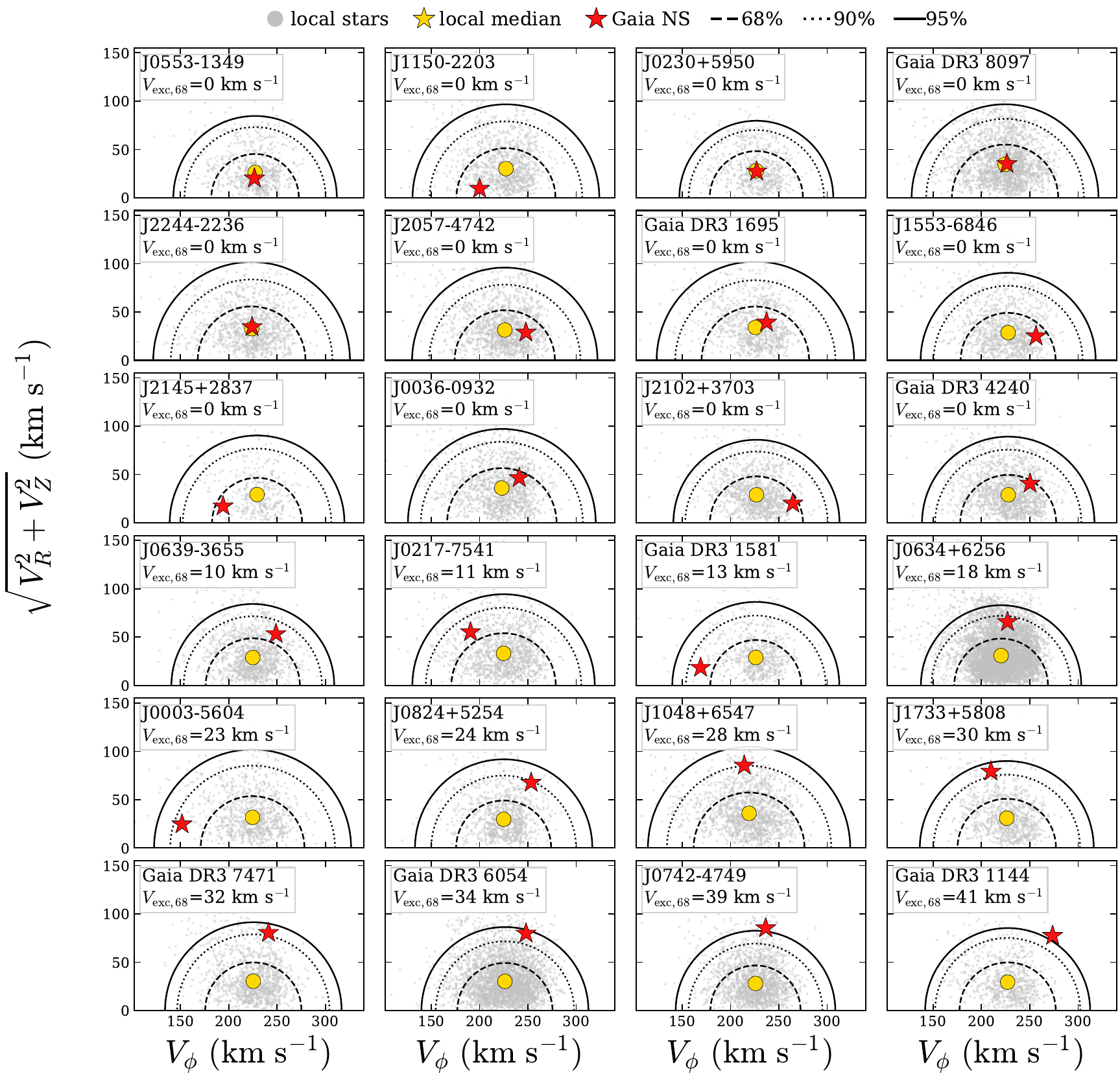}
    \caption{Toomre diagrams for the Gaia NS candidates.  Gray points show the comparison stars for each Gaia NS candidate (red star), and gold stars mark the local sample median.  The dashed, dotted, and solid curves enclose $68\%$, $90\%$, and $95\%$ of the comparison sample. Gaia NS space velocities are similar to, or mildly elevated above, those of neighboring stars, supporting generally low recoil kicks.}
    \label{fig:toomre}
\end{figure*}

This kinematic analysis constrains the systemic recoil, not the natal kick alone. Namely, the binary's space velocity (i.e., center-of-mass motion) depends on mass loss, kick magnitude, kick direction, pre-supernova orbit, and Galactic evolution \citep[e.g.,][for a discussion]{Atri2019}.
In the pre-supernova center-of-mass frame, the systemic recoil of the surviving NS--MS binary is
\begingroup
\begin{equation}
    \mathbf{V}_{\rm sys}
    =
    \frac{
    M_{\rm NS}\mathbf{v}_{\rm kick}
    -
    \Delta M~\mathbf{v}_{\rm He}}
    {M_{\rm NS}+M_2},
    \label{eq:systemic_recoil}
\end{equation}
\endgroup
where $\Delta M=M_{\rm He}-M_{\rm NS}$ and $\mathbf{v}_{\rm He}$ is the pre-supernova orbital velocity vector of the exploding helium star around the binary center of mass.  The first term is the momentum added by the natal kick, while the second term is the Blaauw recoil from symmetric mass loss \citep{Blaauw1961,Hills1983,Kalogera1996,Kalogera1998}.  Applying Equation~\ref{eq:systemic_recoil} to the best-fit distributions of $P_{\rm preSN}$, mass loss, kick speed, and kick direction from our joint fits (Section \ref{sec:joint}) gives a best-fit median $V_{\rm sys}=13~\kms$, with a $1\sigma$ range of $8$--$30~\kms$.

Figure~\ref{fig:toomre} compares each Gaia NS candidate to nearby Gaia DR3 stars on a Toomre diagram $(V_\phi,\sqrt{V_R^2+V_Z^2})$.  Most Gaia NSs have velocities comparable to the dispersion of their local stellar populations.  Of the $24$ non-halo systems, $12$ lie inside the local 68th-percentile contour and therefore have $V_{\rm exc,68}=0$. The median system lies at percentile $0.714$ of its local velocity distribution, and the sample median is $V_{\rm exc,68}=4.9~\kms$. Six systems lie above the local 90th percentile, and two lie above the 95th percentile. For the $12$ systems with $V_{\rm exc,68}>0$, the median excess is $26.2~\kms$, with a central 68\% range of $12.5$--$34.9~\kms$.

\subsubsection{Forward-modeling the local velocity distribution}
\label{subsec:toomre_forward_model}

\begin{figure}
    \centering
    \includegraphics[width=0.99\linewidth]{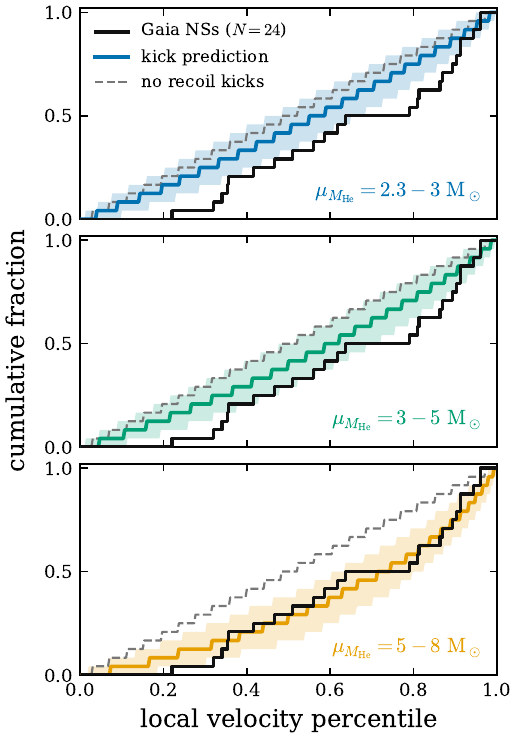}
    \caption{Predicted kinematics of Gaia NSs and their dependence on progenitor mass. 
    We show the distribution of the Gaia NS candidates’ velocity percentiles relative to their local stellar populations (black; see Figure \ref{fig:toomre}). The results are shown after conditioning the posterior on $\mu_{\rm M_{\rm He}} = 2.3$--$3$, $3$--$5$, and $5$--$8~\msun$ (colored curves). Gray dashed curves show the results assuming no recoil kicks. The unified kick model reproduces the observed Gaia NS kinematics if the progenitor mass loss, and hence the Blaauw recoil, increases. 
    }\label{fig:toomre_ppc}
\end{figure}

We forward model the Gaia NS kinematics using the unified kick prescription (Section~\ref{sec:joint}). For each of the $24$ non-halo Gaia NS candidates, we draw kicks from the posterior distributions (Figure~\ref{fig:parent_fourpanel}) and select simulated survivors with similar orbits ($|\Delta\log_{10}P|\leq0.05$ and $|\Delta e|\leq0.05$). The largest $P$ and $e$ uncertainties are $0.0011$ dex and $0.014$ \citep[][]{ElBadry2024b,ElBadry2026}, justifying the chosen range. We add each survivor's recoil (Equation~\ref{eq:systemic_recoil}) to the velocity of a randomly selected nearby field star and compare the resulting local velocity percentile with the observed value.  
We generate $100$ mock $24$-system catalogs for each of $100$ posterior draws of the kick distribution, giving $10{,}000$ catalogs in total.

Figure~\ref{fig:toomre_ppc} compares the predicted and observed local velocity percentiles of the Gaia NS candidates for three progenitor-mass ranges. The $M_{\rm He}=2.3$--$3~\msun$ model predicts a population that is systematically colder than observed, while the $3$--$5~\msun$ prediction lies closer to the data and the $5$--$8~\msun$ prediction broadly reproduces the observed distribution. This trend arises because more massive helium stars eject more mass and produce larger systemic recoils. 
After marginalizing over the full posterior, the predicted median local percentile is $0.60$ ($0.49$--$0.70$), compared with the observed value of $0.71$, with a tail probability of $14\%$. The current kinematics therefore mildly favor greater mass loss but remain consistent with the unified kick model. We discuss the physical implications of this result in Section \ref{subsec:physical_interpretation}.

For comparison, we repeat the calculation without adding a systemic recoil: i.e., the velocity of each Gaia NS candidate is randomly selected from one of its neighboring stars.
This no-kick population is noticeably colder than the Gaia NS observations.
Twelve Gaia NSs reside inside the local $68$th-percentile contour, whereas the no-recoil assumption predicts that $16$ systems ($14$--$19$) should.
Only $2\%$ of no-kick realizations contain $\leq 12$ binaries inside the $68$th-percentile contour.
The Gaia NS population is therefore mildly hotter than its local disk population, as predicted by the kick law derived from their binary orbits.

\subsection{Comparison to NS-LMXBs}\label{subsec:lmxb_compare}

NS-LMXBs are another class of old NS binaries with low-mass companions. Unlike the detached Gaia NSs, the present-day orbits of NS-LMXBs are not very informative for constraining kicks because they experienced significant post-SN orbital evolution through tides, magnetic braking, Roche-lobe overflow, and donor evolution \citep{Webbink1984,Kalogera2000,Ivanova2013, TaurisvandenHeuvel2023}.
Their systemic velocities, however, still retain information about the recoil imparted during core collapse.
Because Gaia NSs and NS-LMXBs are both typically several Gyr old \citep[e.g.,][]{Bahramian2023}, their present-day velocities are also affected by Galactic heating \citep{Wielen1977}. Assuming roughly similar ages, this effect should partly average out in a population-level comparison.

\begin{figure}
    \centering
    \includegraphics[width=0.99\linewidth]{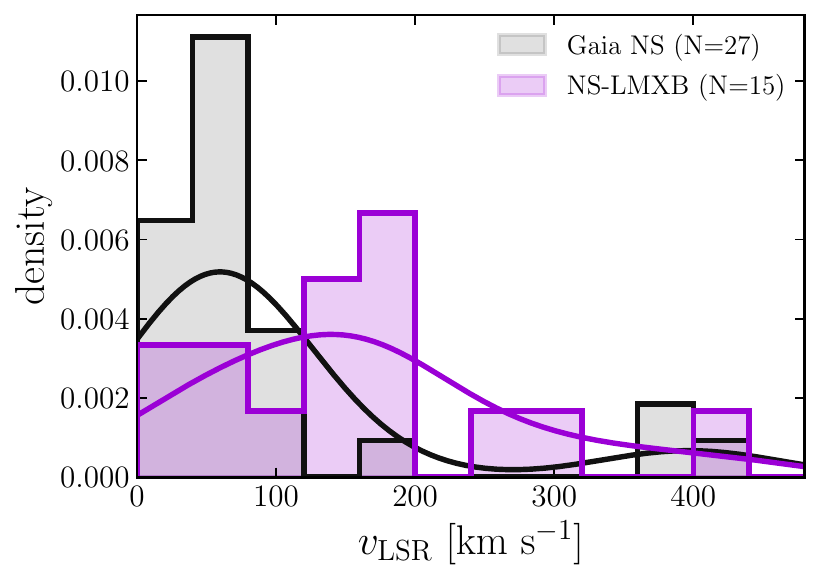}
    \caption{Galactocentric velocity distributions of Gaia NS candidates and NS-LMXBs. NS-LMXBs overlap the Gaia NSs at low velocity but have a more prominent high-velocity tail.}
    \label{fig:lmxb_velocity_compare}
\end{figure}

Figure \ref{fig:lmxb_velocity_compare} compares the present-day systemic velocities of Gaia NSs to NS-LMXBs. The NS-LMXB sample contains $N=15$ systems with well-measured radial velocities, distances, and proper motions \citep{Zhao2023}.
We compare the velocities with respect to the local standard of rest:
\begin{equation}
    v_{\rm LSR}=[V_R^2+V_Z^2+(V_\phi-225~\kms)^2]^{1/2},
\end{equation}
where $225~\kms$ is the adopted local circular speed.

NS LMXBs are kinematically hotter than Gaia NSs.
The Gaia sample has median $v_{\rm LSR}\approx59~\kms$ ($39$--$86~\kms$), while the NS-LMXBs have $v_{\rm LSR}\approx156~\kms$ ($53$--$243~\kms$).
Thus, the typical NS-LMXB is $2$--$3\times$ faster than the typical Gaia NS.
NS-LMXBs also have a larger vertical scale height: Gaia NS candidates have median $|z|=0.37$ kpc ($0.08$--$0.53$ kpc), whereas NS-LMXBs typically extend to $|z|\gtrsim1$ kpc \citep{vanParadijs1995,Jonker2004,ElBadry2024b}.

The hotter kinematics of NS-LMXBs do not necessarily require them to receive larger natal kicks than Gaia NSs, if NS-LMXBs had tighter pre-supernova orbits. 
Indeed, the standard picture of NS-LMXB formation predicts that the main-sequence companion experienced common envelope evolution with the NS progenitor, leading to a very tight orbit prior ($P_{\rm orb}\sim1$--$10$~days) to core collapse \citep[e.g.,][]{Bhattacharya1991,Kalogera1998, Podsiadlowski2002}. 
Gaia NSs, on the other hand, almost surely had pre-SN periods of $P_{\rm orb}=50-500$~days, owing to their present-day low eccentricities \citep[Figure \ref{fig:parent_fourpanel}; see also][]{ElBadry2024d}. 
For a circular pre-supernova orbit, the Blaauw recoil scales as (Equation \ref{eq:systemic_recoil})
\begin{equation}
    V_{\rm Blaauw}\propto v_{\rm He}\propto a^{-1/2}\propto P_{\rm orb}^{-1/3}.
\end{equation}
Thus, if LMXBs have $\sim10$--$50\times$ {\it smaller} initial orbital periods ($P_{\rm orb}$) than Gaia NSs, they should receive $\sim2.2$--$3.7\times$ {\it larger} Blaauw recoils at fixed mass loss and kick velocity. 

We test the combined effects of natal kicks, mass loss, and binary survival by applying the unified kick prescription to a toy population of circular LMXB progenitors with $M_2=1~\msun$, $M_{\rm He}=3.6\pm1.2~\msun$, and $0.1<P_{\rm preSN}/{\rm d}<1000$. Bound systems with $1<r_{\rm peri}/\rsun<10$ have a median birth recoil of $158~\kms$, with a central $68\%$ range of $86$--$258~\kms$. 
Selecting systems with $P_{\rm postSN}<10$ days gives a similar median of $163~\kms$ ($89$--$263~\kms$). Both predictions agree with the NS-LMXB velocity distribution \citep{ODoherty2023}, which has a median of $136~\kms$ and a central $68\%$ range of $79$--$233~\kms$. In our model, $80\%$ of these LMXBs arise from the high-kick component.
The same parent kick law (Figure \ref{fig:parent_kick_law}) can therefore explain both populations: compact LMXB progenitors receive larger Blaauw recoils and mainly arise from the high-kick branch, whereas wide Gaia NS binaries preferentially arise from the low-kick branch.

\subsection{Be+NS peculiar velocities}\label{subsec:bens_kinematics}

The unified model predicts that the low-eccentricity Be+NS systems receive smaller typical recoils than the high-eccentricity systems.
We test this prediction using the $23$ Be+NS binaries compiled by
\citet{Valli2025}. Twenty have Gaia DR3 counterparts, $19$ have 5-
or 6-parameter astrometric solutions, and $13$ satisfy
$\varpi/\sigma_\varpi\geq4$. The final sample contains $3$ low-$e$
systems, defined as $e\leq0.25$, and $10$ high-$e$ systems, defined as
$e>0.25$ \citep{Valli2025}.
Among these systems, $5$ have published binary center-of-mass radial velocities and therefore direct 3D peculiar velocities.  For the remaining eight, we follow \citet{Nuchvanichakul2025} in adopting the isotropy-scaled estimate $V_{\rm pec}^\dagger=(4/\pi)V_{\rm pec, 2D}$. The source measurements and references are given
in Appendix~\ref{subapp:toomre_bens} and Table~\ref{tab:bens_kinematics}.

\begin{figure}
    \centering
    \includegraphics[width=0.9\columnwidth]{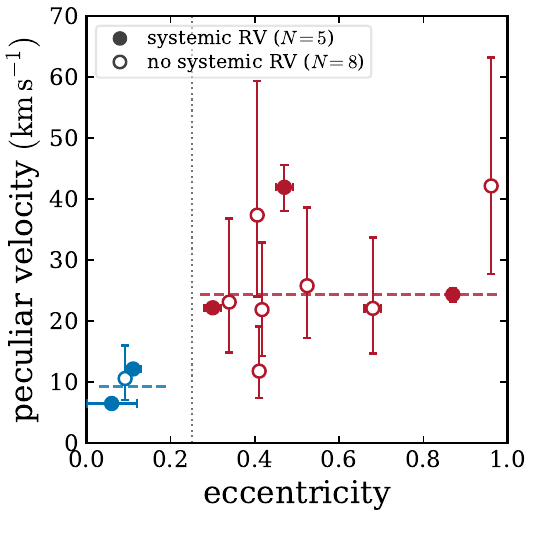}
    \caption{ Peculiar velocity as a function of orbital eccentricity for the Be+NS binaries.
    Filled points have full 3D peculiar velocities. Open points have only 2D constraints, so the inferred peculiar velocities marginalize over the line-of-sight component. The vertical dotted line shows $e=0.25$, the high/low boundary defined by \citet{Valli2025}.
    Vertical error bars show 16th--84th percentile velocity intervals. Horizontal dashed lines mark the median systemic velocities for low- and high-$e$ binaries. High-$e$ systems have systematically larger peculiar velocities than low-$e$ systems, consistent with the high-$e$ branch receiving larger natal kicks.}
    \label{fig:bens_kinematics}
\end{figure}

Figure~\ref{fig:bens_kinematics} compares the present-day peculiar velocities of Be+NS binaries as a function of their eccentricity.
Among the five systems with measured 3D velocities, the two low-$e$ binaries have $V_{\rm pec}=6.4$ and $12.1~\kms$, while the three high-$e$ binaries have $V_{\rm pec}=22.2$, $24.3$, and $41.9~\kms$.  An exact two-sided permutation test on the difference in mean velocity gives $p=0.10$.  Including the eight systems without systemic RVs gives median velocities of $10.6$ and $23.7~\kms$ for the low- and high-$e$ branches, respectively, with $p=0.010$ for the full $N=13$ sample.  This test uses the central velocity estimates and does not propagate the uncertainty introduced by converting projected velocities to 3D speeds.  The data therefore suggest that high-$e$ Be+NS binaries are kinematically hotter, although additional systemic RV measurements and a larger sample size are needed to establish the difference.

To compare this trend with the unified kick model, we draw progenitors and kicks from the joint posterior (Figure \ref{fig:parent_fourpanel}), apply the supernova transformation, and retain bound systems in the observable Be+NS period--eccentricity region.  For each observed Be+NS binary, we use its measured orbital parameters and select model survivors within $|\Delta\log_{10}P|\leq0.025$ and $|\Delta e|\leq0.020$. This produces simulated catalogs of $N=5$ Be+NS binaries whose orbits match the observed systems with measured 3D velocities. 
The model predicts a high-$e$/low-$e$ systemic velocity ratio of $2.25$ ($1.69$--$3.07$) and a difference of $10.8~\kms$ ($6.4$--$16.9~\kms$). Matching all $13$ systems with at least 2D measurements gives a ratio of $1.88$ ($1.50$--$2.34$) and a difference of $9.0~\kms$ ($5.9$--$12.7~\kms$), with a $99.3\%$ posterior-predictive probability that the high-$e$ mean is larger.
The concordance model therefore predicts the trend seen in Figure~\ref{fig:bens_kinematics}: high-$e$ Be+NS binaries should move $\approx2\times$ faster than low-$e$ systems. 

A model predicting Be+NS binaries from only low kicks \citep[e.g.,][]{DisbergMandelHirai2026} would not generally predict a space velocity difference between low- and high-$e$ binaries. 
The observed velocity difference provides tentative support for larger kicks among high-$e$ binaries, although mass loss and pre-supernova motion could also contribute (Section~\ref{subsubsec:comp_to_BeNS}).

\section{Survival of Tertiaries}\label{sec:tertiaries}

\begin{figure*}
    \centering
    \includegraphics[width=0.99\linewidth]{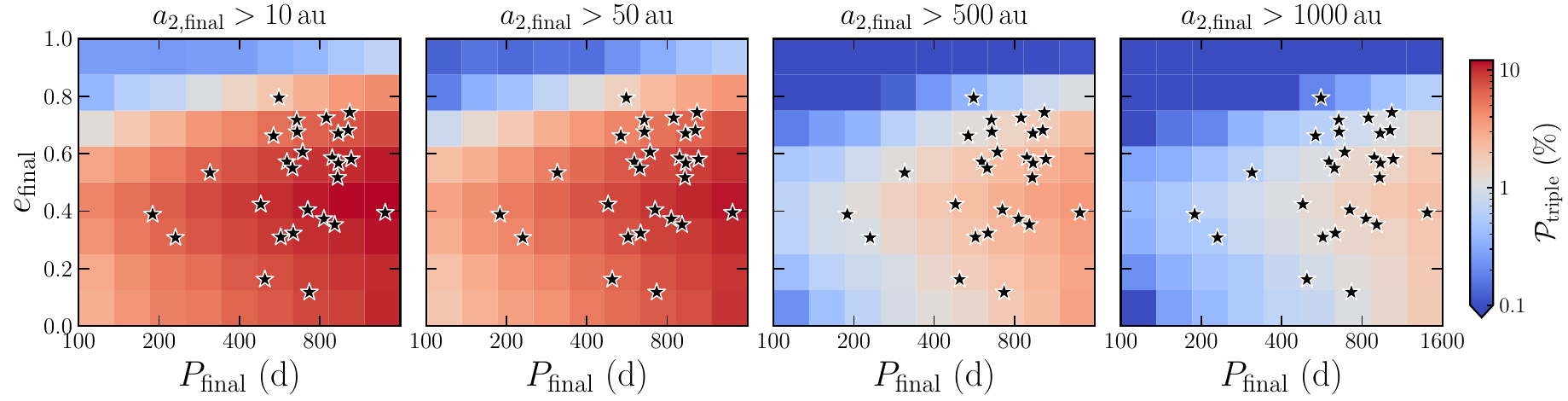}
    \caption{Survival of tertiary companions to Gaia NS binaries. 
    Each cell shows the percentage of Gaia NS inner binaries in that period--eccentricity bin that are predicted to retain a stable tertiary with $a_{2,{\rm final}}>10$, $50$, $500$, or $1000$~au.
    Gaia NSs with long orbital periods are more likely to harbor bound tertiaries, since these binaries generally received the weakest kicks.
    Increasing the tertiary separation selects Gaia NSs with progressively lower center-of-mass recoils. }
    \label{fig:tertiary_survival}
\end{figure*}

The progenitors of NSs, massive OB stars ($\gtrsim10~{\rm M_\odot}$), are often born in triple or higher-order multiples \citep[$50$--$80\%$;][]{Moe2017,Offner2023}. This suggests that tertiary companions are prevalent around the progenitors of Gaia NSs. 
Triple evolution can also play a role in the formation of NSs and BHs orbited by low-mass companions \citep[e.g.,][]{Eggleton1986,Naoz2016a,Rose2019, Generozov2024,Li2024,Shariat2025b,Li2026,Naoz2025}.
A tertiary that remains bound after NS formation provides an independent constraint on the center-of-mass recoil, as illustrated by the wide companion to the BH-LMXB V404 Cygni \citep{Burdge2024,Shariat2025b}
Motivated by these results, we investigate how the inferred natal kicks impact tertiary companions.

\subsection{Tertiary-survival calculation}

We add outer companions to simulated Gaia NS progenitors and determine whether the resulting triples remain bound and dynamically stable after the supernova. The tertiary properties are drawn from the \citet{Shariat2025d} triple model, which was calibrated on resolved triples in {\it Gaia} DR3. We select systems with NS-forming primaries, $8\leq M_1/\msun\leq20$, low-mass inner companions, $q_{\rm in}=M_2/M_1<0.2$, and inner separations $1<a_1/{\rm au}<10$. The tertiary mass, outer semimajor axis, eccentricity, and mutual inclination define the immediate pre-supernova outer orbit.

Our fiducial calculation uses the orbital period distribution of \citet{Duquennoy1991}, as adopted by \citet{Shariat2025d}. Although this distribution was measured primarily for lower-mass stars, it is broadly consistent with recent constraints on massive hierarchical triples \citep{Bordier2026}; uniform and log-uniform outer-separation distributions give similar results. We assign one tertiary to each progenitor drawn from the inferred Gaia NS kick model and retain systems that are initially hierarchical \citep{Naoz2013} and satisfy the \citet{Mardling2001} stability criterion. {We then apply the instantaneous-supernova calculation described in Appendix \ref{app:sn_formalism} \citep{Lu2019}.} A tertiary survives if the inner binary becomes Gaia-NS-like and the post-supernova triple remains bound, hierarchical, and dynamically stable. We simulate $5\times10^6$ initially stable triples. This calculation treats the assigned outer orbit as the immediate pre-supernova configuration and does not follow its earlier evolution.

\subsection{Results}

\vspace{-0.8cm}
\begin{deluxetable*}{llrcccc}
\tablecaption{Survival of wide tertiaries around Gaia NS and Be+NS binaries.\label{tab:tertiary_summary}}
\tablehead{
\colhead{Sample} & \colhead{Orbit} & \colhead{$N$} &
\colhead{$M_3$} & \colhead{$a_2$} & \colhead{$e_2$} &
\colhead{$i_{\rm mut}$} \\
\colhead{} & \colhead{} & \colhead{} &
\colhead{($\msun$)} & \colhead{(au)} & \colhead{} & \colhead{(deg)}
}
\startdata
\sidehead{\textbf{Gaia NS binaries}}
initially stable triples & pre-SN & $5.00\times10^{6}$ &  $4.8^{+15.8}_{-2.9}$ &  $693^{+6991}_{-596}$ & $0.63^{+0.23}_{-0.29}$ &  $90.6^{+43.0}_{-43.4}$ \\
stable survivors & pre-SN &  $3.30\times10^{5}$ &  $11.3^{+27.4}_{-8.4}$ &  $101^{+209}_{-65}$ &  $0.52^{+0.23}_{-0.24}$ &  $91.8^{+40.9}_{-41.3}$ \\
stable survivors & post-SN &  $3.30\times10^{5}$ &  $11.3^{+27.4}_{-8.4}$ &  $164^{+617}_{-116}$ &  $0.62^{+0.24}_{-0.29}$ &  $92.4^{+42.6}_{-43.9}$ \\
\sidehead{\textbf{Be+NS binaries}}
initially stable triples & pre-SN & $5.00\times10^{6}$ & $9.8^{+5.5}_{-5.4}$ &  $680^{+6838}_{-585}$ &  $0.64^{+0.22}_{-0.29}$ &  $90.5^{+42.9}_{-43.3}$ \\
stable survivors & pre-SN &  $9.66\times10^{5}$ & $10.5^{+5.1}_{-5.7}$ &  $101^{+222}_{-68}$ &  $0.53^{+0.24}_{-0.25}$ &  $90.6^{+42.7}_{-41.8}$ \\
stable survivors & post-SN &  $9.66\times10^{5}$ & $10.5^{+5.1}_{-5.7}$ &  $144^{+586}_{-106}$ &  $0.64^{+0.23}_{-0.30}$ &  $91.1^{+43.1}_{-43.4}$ \\
\enddata
\tablecomments{Reported values are medians with 16th--84th percentile uncertainties. The survivor samples are conditioned on the inner binary becoming either Gaia NS-like or Be+NS-like and the outer orbit remaining bound and dynamically
stable after the supernova.}
\end{deluxetable*}

Table~\ref{tab:tertiary_summary} summarizes the initial tertiary population and the subset that survives the supernova.  
Surviving tertiaries are preferentially those that are initially massive with tighter outer orbits around the Gaia NS progenitor. 
The stable survivors have median initial $M_3=11.3~\msun$ and $a_{2,i}=101$ au, compared to $M_3=4.8~\msun$ and $a_{2,i}=693$ au for the entire initially stable population. The outer orbit then expands and becomes more eccentric during the supernova, with median post-supernova values $a_{2,f}=164$ au and $e_{2,f}=0.62$, reflecting the mass loss and center-of-mass recoil of the inner binary.

Figure~\ref{fig:tertiary_survival} shows the conditional probability that a Gaia NS inner binary retains a bound and stable tertiary,
\begin{equation}
\begin{aligned}
    \mathcal{P}_{\rm triple}
    = 
    \frac{
    N_{\rm tertiary}\left(P_{\rm final},e_{\rm final},a_{2,\rm final}>a_{\rm cut}\right)
    }{
    N_{\rm GaiaNS}\left(P_{\rm final},e_{\rm final}\right)
    } .
    \label{eq:p_triple}
\end{aligned}
\end{equation}
Here $N_{\rm GaiaNS}(P_{\rm final},e_{\rm final})$ is the number of simulated systems whose inner binary becomes Gaia-NS-like in a given period--eccentricity bin, and $N_{\rm tertiary}$ is the subset of those systems that also retain a bound, dynamically stable tertiary with final outer semimajor axis larger than $a_{\rm cut}$. 
The four panels show this probability for $a_{\rm cut}=10$, $50$, $500$, and $1000$ au. 
Across the full Gaia NS inner-binary sample, the retention fractions are $6.6\%$, $5.5\%$, $1.5\%$, and $0.86\%$ for these four cuts, respectively. 
{Retention is highest for long-period Gaia NS inner binaries, whose orbits imply smaller systemic recoils. 
Low eccentricity does not necessarily imply a small recoil: a kick opposite the progenitor's orbital motion can reduce the inner eccentricity while adding to the recoil from supernova mass loss. The widest tertiaries are especially sensitive to this recoil, so their survival favors long-period inner binaries with intermediate eccentricities rather than the most circular systems.}

Wide tertiaries are more easily disrupted because their orbital velocities decrease as $v_{\rm orb}\propto (GM_{\rm tot}/a_2)^{1/2}$. For a $1~\msun$ tertiary, $v_{\rm orb}$ decreases from approximately $17~\kms$ at $a_2=10$ au to $1.7~\kms$ at $1000$ au. Wide tertiaries are therefore unbound by even modest center-of-mass recoils\footnote{The relevant ``kick'' seen by the tertiary is the recoil of the inner binary, not just the natal kick velocity of the NS, although the two are related through Equation~\ref{eq:systemic_recoil}.} at wide separations.

\subsection{Be+NS binaries}

\begin{figure*}
    \centering
    \includegraphics[width=0.99\linewidth]{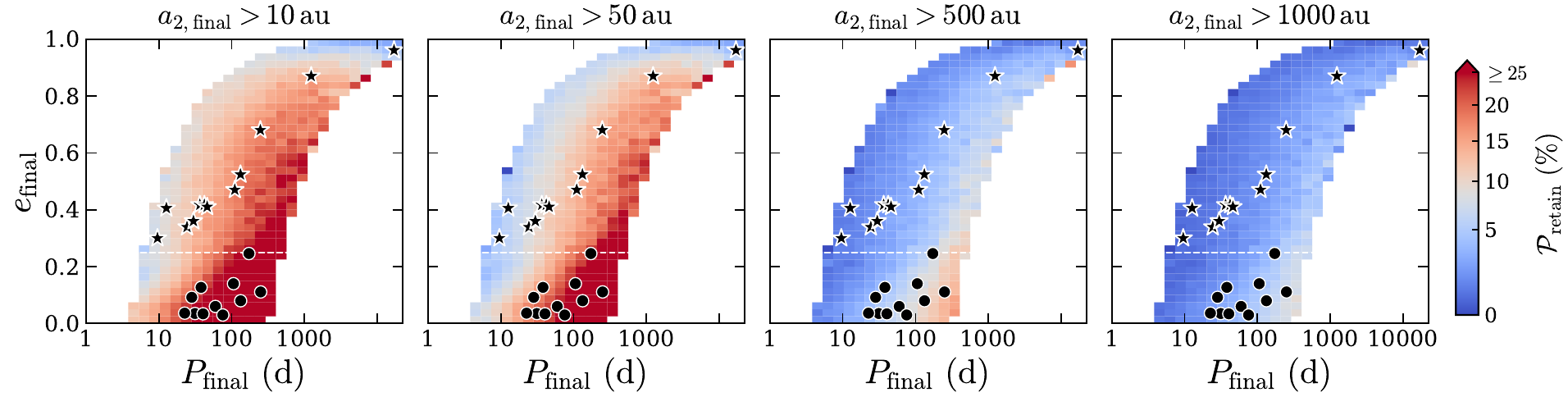}
    \caption{Survival of tertiary companions to Be+NS binaries. Each cell gives the percentage of selected Be+NS inner binaries in that period--eccentricity bin that retain a bound, dynamically stable tertiary with $a_{2,{\rm final}}>10$, $50$, $500$, or $1000$~au. Black circles and stars mark the observed low- and high-eccentricity systems, with the dividing line at $e=0.25$. Low-eccentricity and long-period Be+NS binaries retain tertiaries more often because they generally received lower kicks.}
    \label{fig:bens_tertiary_survival}
\end{figure*}

We repeat the tertiary-survival calculation using Be+NS progenitors and kicks drawn from the joint posterior. We fix the Be-star mass to $15~\msun$ and apply the same selections used in the joint likelihood. The outer separation, eccentricity, and mutual inclination follow the fiducial distributions used for the Gaia NS calculation, while the tertiary mass ratio is drawn uniformly over $0.1<q_{\rm out}<1$. The calculation draws $5\times10^6$ initially stable triples.

Figure~\ref{fig:bens_tertiary_survival} shows that $19.1\%$, $15.1\%$, $4.2\%$, and $2.4\%$ of Be+NS binaries retain stable tertiaries beyond $10$, $50$, $500$, and $1000$~au, respectively. For $a_{2,{\rm final}}>50$~au, the retention fraction is $20.7\%$ for low-eccentricity systems and $10.6\%$ for high-eccentricity systems. Using the same tertiary distributions, Be+NS binaries retain tertiaries $\sim3\times$ more often than Gaia NS binaries because their more massive inner binaries are harder to unbind. The surviving tertiary companions have a median mass of $M_3=10.5^{+5.1}_{-5.7}~\msun$ (Table~\ref{tab:tertiary_summary}).

\subsection{Observed NSs in triples}
Finding NSs in hierarchical triples can test the proposed kick model.  Recently, PSR J0435+3233, an 8-day pulsar--white-dwarf binary, was found to have a $\sim1~\msun$ tertiary with $a_2\approx25$ au and $e_2\approx0.6$ \citep{Freire2026,Yang2026}.  Applying our universal kick model to the pre-supernova configuration adopted by \citet{Yang2026}, we find that $22$--$25\%$ remain bound and dynamically stable for the low-kick component inferred in this work, compared with only $\approx4\%$ under the all-NS kick law.  
The more compact triple PSR J0337+1715 has an outer white dwarf in a 327-day orbit and may have survived a $\sim110$--$125~\kms$ kick \citep{Ransom2014,Voisin2025}. 
Applying our model to its immediate pre-supernova configuration gives stable-survival fractions of $\approx95\%$ for the low-kick component and $\approx20\%$ for the all-NS law. 
Thus, both branches of our unified kick model can produce surviving triples, but $\sim100~\kms$ kicks require more tightly bound outer orbits.

\citet{Nagarajan2026ThirdWheel} searched $12$ Gaia NSs for tertiary companions with Keck/NIRC2 and identified one viable candidate near J1733+5808. The candidate is an approximately $0.23~\msun$ M dwarf at a projected separation of $\sim3000$ au, with a $1.6\%$ chance-alignment probability. It is not detected in {\it Gaia} DR3, so common proper motion has not yet been established. Conditioning our calculation on the period and eccentricity of J1733+5808 and fixing $M_3=0.25~\msun$, we find an approximately $0.15\%$ probability of retaining a stable tertiary with $a_{2,\rm final}>1000$ au. The inferred kick distribution therefore allows such a companion, but predicts that it should be rare.

The apparent paucity of tertiaries around Gaia NSs may partly be driven by survey selection effects and astrophysical biases.  
The {\it Gaia} astrometric binary catalog, from which Gaia NS candidates were selected, excludes sources with ${\tt ipd\_frac\_multi\_peak} > 2$ and 
${\tt ipd\_gof\_harmonic\_amplitude} \geq 0.1$ \citep{Halbwachs2023}.
These cuts remove binaries whose {\it Gaia} windows show multiple image peaks or scan-angle-dependent image-fit residuals, so they disfavor systems with luminous tertiaries at marginally resolved separations \citep[$0.1$--$1\arcsec$;][their figure 6]{Tokovinin2023}. Such companions can also dilute or bias the photocenter motion of the inner NS--MS binary, in some cases affecting the astrometric signal enough that systems are missed by {\it Gaia}'s non-single-star catalog.
This bias is relevant because most surviving tertiaries in our model are massive ($\gtrsim3~\msun$) and within $\sim1000$ au (Table~\ref{tab:tertiary_summary}). {Most of these tertiaries would become compact remnants (white dwarfs, neutron stars, or black holes) within $\lesssim1$ Gyr. These compact objects would generally be too faint for existing imaging surveys, while others may have been unbound by subsequent supernovae.} Dedicated searches with {\it Gaia} DR4 and {\it Roman} would be highly instructive.

\section{Discussion}\label{sec:discussion}

\subsection{Comparison to Previous Kick Constraints}\label{subsec:previous_kick_constraints}

\begin{figure}
    \centering
    \includegraphics[width=0.99\columnwidth]{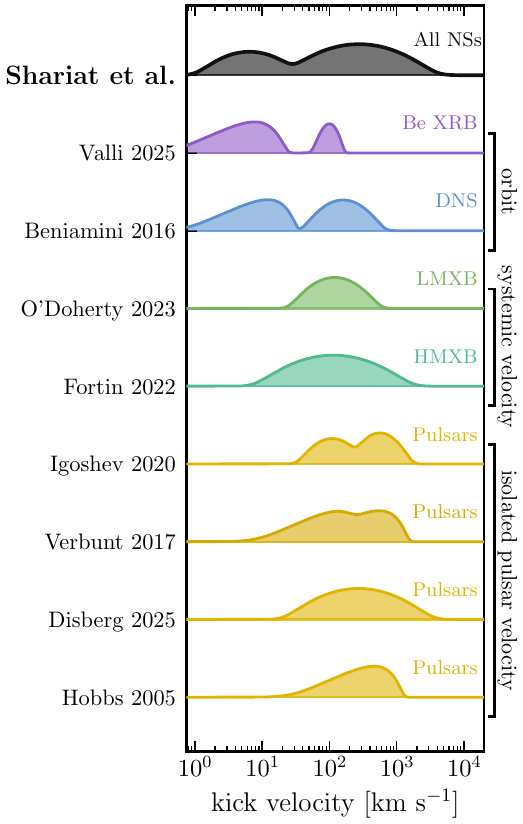}
    \caption{Empirical neutron star kick constraints.  Each curve is normalized to its own maximum with the vertical axis log-scaled. The rows are grouped by the observable used for inference: post-supernova orbits, surviving-binary systemic velocities, or isolated-pulsar space velocities (figure adapted from \citealt{Valli2025}).  
    The top curve (black) shows the parent kick distribution for all NSs inferred in this work.}
    \label{fig:previous_kick_constraints}
\end{figure}

Figure~\ref{fig:previous_kick_constraints} compares different empirically inferred kick distributions, which fall into two broad categories: orbital constraints and peculiar velocity constraints. For Gaia NSs, wide Be XRBs, and DNSs, the kick is inferred mainly from the post-supernova orbit \citep{Beniamini2016,Valli2025}. For NS-LMXBs, NS-HMXBs, and isolated pulsars, the constraints are primarily inferred from present-day space velocities and Galactic kinematics \citep{Hobbs2005,FaucherGiguere2006,Verbunt2017,Igoshev2020,Disberg2025}. 

{\citet{DisbergMandelHirai2026} instead test five fixed kick prescriptions with \texttt{COMPAS} \citep{Riley2022,Mandel2025}, evolving $10^6$ solar-metallicity binaries and comparing Gaia NSs, NS-LMXBs, NS--WD binaries, Be XRBs, and DNSs. Their calculation follows mass transfer, common-envelope evolution, eccentric pre-supernova orbits, two supernovae for DNSs, and delayed rocket kicks. Given the uncertainties in binary evolution, particularly for Gaia NSs, we adopt an empirical approach. We infer the shared kick law and progenitors from the observed binary orbits, birth rates, and young-pulsar velocities while remaining agnostic to the binary evolution history.}

\subsubsection{Gaia NS binaries}\label{subsubsec:comp_to_GaiaNS}

{\citet{DisbergMandelHirai2026} use the original $21$ Gaia NS candidates \citep{ElBadry2024b} and an analytic approximation of the Gaia selection function \citep{Baibhav2026}. We use the updated $27$-candidate sample \citep{ElBadry2026}, the full DR3 astrometric binary selection function, and the intrinsic Galactic birth rate inferred by \citet{Shariat2026GaiaNSDemographics} to connect the observed binaries to NS birth kicks.
With their default common-envelope model, pulsar-like natal kicks ($\gtrsim100~\kms$) followed by a fixed $30~\kms$ rocket kick \citep{Hirai2024} best reproduce the Gaia NS population. Their two-stage common-envelope model permits low natal kicks only with a rocket kick; otherwise the binaries are too eccentric. This rocket kick model, they note, predicts kinematic velocities that are slower than those observed for Gaia NSs, unless additional heating is invoked. The physical feasibility of these large and sustained rocket kicks remains debated (\citealt{Baibhav2026}; Section~\ref{subsec:rocket_kicks}).
}

{Our inferred au-scale progenitors reproduce the Gaia NS orbits without rockets, predominantly with kicks $\lesssim50~\kms$. Larger ejecta masses ($M_{\rm He}=4$--$6~\msun$) predict systemic velocities in agreement with those observed for Gaia NSs (Figure~\ref{fig:toomre_ppc}). 
We do not follow the preceding mass transfer history and therefore cannot determine how Gaia NSs remained wide or whether they experienced common-envelope evolution. 
However, {\it Gaia} DR3 has revealed WD+MS binaries in au-scale post-interaction orbits \citep[e.g.,][]{Shahaf2024,Yamaguchi2024,Yamaguchi2024b,Yamaguchi2025, vanSon2026}, and 
\citet{Shariat2026GaiaNSDemographics} show that the Gaia NS and WD+MS populations have similar period and MS-star mass distributions. These WD+MS binaries show that binary interaction can leave compact remnants in similarly wide orbits. 

}

\subsubsection{Be+NS binaries}\label{subsubsec:comp_to_BeNS}

{Wide, low-eccentricity Be XRBs provided early evidence for low natal kicks \citep{Pfahl2002,Podsiadlowski2004}.} Recently, \citet{Valli2025} modeled the low-$e$ and high-$e$ Be+NS binaries separately, finding ultralow kicks ($\lesssim10~\kms$) and low mass loss ($\Delta M\lesssim1$--$2~\msun$) for the low-eccentricity systems, and $100~\kms$ with $\Delta M\sim3~\msun$ for the high-eccentricity systems.
The latter branch favored polar-aligned kick directions, which was also previously invoked for some young pulsars from spin--velocity alignments \citep[e.g.,][]{Kramer2003,Johnston2005,Chatterjee2009,Noutsos2012,Noutsos2013,Mandel2023}.  
Our unified fit also recovers an ultralow component with median $6.5~\kms$ for Be+NS binaries, while the low-velocity tail of the high-kick component supplies the eccentric Be+NS systems near $100~\kms$.  The present data therefore do not \emph{strictly require} a separate intermediate-velocity parent component.

After performing various tests with polar kicks, we find that a narrowly polar-kick prescription applied to all Gaia NS progenitors is disfavored.
Such restrictive geometry fails to produce any low-$e$ ($e < 0.35$) systems, inconsistent with the $5$ observed. These instead require a kick direction that partly cancels out the original orbital motion (Section~\ref{sec:gaia_only})
This does not exclude the $\sim100~\kms$ polar component inferred by \citet{Valli2025}.  Kicks of this magnitude rarely leave wide Gaia NS binaries bound, so the current Gaia NS sample provides little constraint on their directional distribution.

This result is consistent with \citet{Baibhav2026}, who found that polar kicks fail for initially circular Gaia NS-like progenitors unless the supernova mass loss is extremely small ($\lesssim1~\msun$). 
Their successful non-isotropic models require much broader half-opening angles of $\gtrsim45^\circ$ for low-mass progenitors and $\gtrsim70^\circ$ for high-mass progenitors.

The inference of polar kicks from \citet{Valli2025} is conditional on how the $23$ Be-XRBs are selected and divided into low- and high-eccentricity groups.  \citet{Valli2025} retain only systems with $30<a_{\rm post}(1-e_{\rm post})/\rsun<450$.  For their $15+1.4~\msun$ reference binary, the lower periastron limit requires $e<0.40$, $0.62$, and $0.79$ at $P_{\rm orb}=10$, $20$, and $50$ d, respectively, and therefore restricts short-period, high-eccentricity binaries from entering the sample.  
The preference for a $\sim5^\circ$ polar cone is driven by fitting the narrow high-$e$ sequence after this periastron cut.

However, Be XRBs can exist in these short-period, high-eccentricity orbits that are not considered in the \citet{Valli2025} fit.  The Galactic gamma-ray pulsar binary LS~I~$+61^\circ303$ has $P_{\rm orb}=26.496$ d and $e=0.537\pm0.034$, where radio pulsations establish that its compact object is an NS \citep{Aragona2009,Weng2022}. \citet{Valli2025} restrict their study to X-ray accretion-powered binaries, which excludes this source from their fit.   
The LMC Be-XRB A0538--66 has $P_{\rm orb}=16.6409$ d and $e=0.72\pm0.14$ \citep{Rajoelimanana2017}, within the `missing' $P$--$e$ region, and is excluded from their fit of only Galactic sources.  They show, however, that the deficit is a property of the selected Galactic accreting Be-XRB sample, not an empirically empty part of the full Be--NS period--eccentricity plane.

Our fitting procedure differs from \citet{Valli2025} in three main ways: (1) we fit the Gaia NS, Be+NS, and isolated pulsar samples simultaneously, (2) we use a single-peaked progenitor-mass distribution for each population rather than a double-peaked $M_{\rm He}$ distribution, and (3) we do not separately fit the low- and high-$e$ Be+NS subsamples. 
The latter two choices keep the model parametrization deliberately simple, given the combined Gaia+Be sample contains only $50$ systems, while still reproducing the observed period--eccentricity distributions reasonably well (Figure~\ref{fig:joint_pe_check}). 
These choices are not demonstrably better than those of \citet{Valli2025}, but address a different question: whether a simpler, shared kick law can reproduce Gaia NSs, Be+NS binaries, and isolated pulsars simultaneously.

With this setup, $\approx90\%$ of selected Gaia NS binaries
and $\approx50\%$ of Be+NS binaries 
are drawn from the low-kick component (Figure~\ref{fig:parent_fourpanel}), the latter being consistent with \citet{Valli2025}.  
In contrast, Gaia NSs fundamentally require a broad $v_{\rm kick}$ range to explain the diversity of orbital configurations, but do not uniquely require either a bimodal parent law or polar kick geometry.
{The inferred Be+NS progenitor periods (Section~\ref{subsubsec:joint_progenitors}) are consistent with \citet{Valli2025}, who found characteristic periods of $\log(P_{\rm preSN}/{\rm d})\sim2$ for the low-$e$ binaries and $\sim1.6$ for the high-$e$ sequence, with a wide spread ($\log(P_\sigma/{\rm day})\sim1$--$2$). Periods of $\sim5$ to several hundred days are also expected for the nearly circular Be + stripped star binaries produced by stable mass transfer \citep{Packet1981,Pols1991,Reig2011,Vinciguerra2020,Liu2024,Valli2025,LopezOller2026}.}

{Both our study and \citet{DisbergMandelHirai2026} use the same $23$ Be+NS binaries and find low kicks ($\lesssim10~\kms$) for the low-$e$ binaries. None of their five kick prescriptions reproduces the high-$e$ sequence from their \texttt{COMPAS} progenitors. They suggest that the high-$e$ sequence could be produced by initially eccentric binaries, which might arise through mergers in triples. Our circular-progenitor model instead produces most high-$e$ systems through $\sim100~\kms$ kicks \citep[see also][]{Valli2025}. 
It therefore predicts that the high-$e$ binaries should have larger systemic velocities than the low-$e$ binaries. Interestingly, the observed high-$e$ Be+NS binaries are indeed faster than the low-$e$ binaries (Figure~\ref{fig:bens_kinematics}). While this provides tentative evidence for the high-$e$ Be+NS binaries receiving larger systemic recoil kicks, the samples remain too small to determine whether the difference is caused by natal kicks, mass-loss recoil, or pre-supernova motion.
}

\subsubsection{Double neutron stars}\label{subsubsec:comp_to_DNS}

\citet{Beniamini2016} modeled 10 Galactic DNS systems and found that the low-eccentricity systems require small mass loss and weak kicks ($\Delta M\lesssim0.5~\msun$ and $v_{\rm kick}\lesssim30~\kms$), while the higher-eccentricity systems allow $\Delta M\sim2.2~\msun$ and kicks up to $\sim400~\kms$.  \citet{Disberg2024} find a similar range for DNS natal kicks ($\sim50~\kms$) from their Galactic kinematics.
The low-eccentricity DNS branch overlaps with the Gaia NS kicks, $v_{\rm kick}=10$--$50~\kms$ (Section~\ref{sec:gaia_only}), but the evolutionary histories are likely different. Gaia NS progenitors had to remain relatively wide before the supernova, whereas DNSs are thought to have descended from a compact, post-interaction NS+He binary \citep[e.g.,][]{Tauris2017, Sgalletta2023}.  

{\citet{Grichener2026} showed that the apparent gap in DNS eccentricities does not necessarily require distinct kick components. With weak kicks from the second supernova, a nonmonotonic relation between progenitor and remnant masses can produce two eccentricity branches through different amounts of supernova mass loss. However, they also note that current DNS sample is also statistically consistent with a smooth eccentricity distribution. \citet{DisbergMandelHirai2026} also reproduce two branches through mass loss with weak second-supernova kicks, consistent with a low-kick component. Some DNSs require at least one large kicks: the spin--orbit misalignment and kinematics of PSR B1534+12 indicate a substantial natal kick \citep[see also][]{Grichener2026}. Although DNSs are not included in our inference, evidence for both weak and strong kicks is qualitatively consistent with our bimodal model.
}

\subsubsection{X-ray binaries}\label{subsubsec:comp_to_XRBs}

The high systemic velocities of NS-LMXBs \citep{ODoherty2023,Zhao2023} are consistent with the bimodal kick law acting on compact progenitors (Section~\ref{subsec:lmxb_compare}).
{\citet{DisbergMandelHirai2026} similarly favor pulsar-like natal kicks because low-kick models underpredict the NS-LMXB systemic velocities. In our model, $80\%$ of NS-LMXBs arise from the pulsar-kick component.}

\citet{Fortin2022} inferred natal kicks for $35$ NS HMXBs using Gaia EDR3 kinematics and orbital information, finding a Gamma-distribution mean of $116^{+18}_{-15}~\kms$ and that Be HMXBs are colder than supergiant HMXBs. 
These estimates are not based on a fully homogeneous or selection-corrected sample: Gaia parallax-quality cuts can remove distant systems, systemic radial velocities are missing for $\sim60\%$ of sources, and the inferred kicks can depend on companion masses and post-supernova binary evolution. 
Nevertheless, the kick scale is broadly consistent with the high-kick branch in our joint fit, $v_{\rm kick}\sim100~\kms$ (Figure~\ref{fig:parent_fourpanel}).

\subsubsection{Isolated Pulsars}\label{subsubsec:comp_to_isolated}

For isolated pulsars, the standard \citet{Hobbs2005} Maxwellian has one-dimensional dispersion $\sigma=265~\kms$ and mean three-dimensional birth speed $\approx400~\kms$.  \citet{Verbunt2017} favored a two-Maxwellian model with mean speeds $\approx120$ and $540~\kms$, and \citet{Igoshev2020} found a young-pulsar bimodal fit with only $5\pm3\%$ of systems below $60~\kms$. The low-velocity mode of these bimodal kick distributions overlaps with the high-velocity branch of the joint-fit distributions from Section \ref{sec:joint}.
Recently, \citet{Disberg2025} argued that the current sample of young isolated pulsars is statistically compatible with a {\it single} lognormal distribution peaking near $\sim150$--$200~\kms$.

We test (Section~\ref{subsec:broad_unimodal_main}) whether the Gaia NSs can be drawn from the low-velocity tail of a single-lognormal distribution inferred for pulsars; they cannot. 
The \citet{Disberg2025} young-pulsar lognormal places only $0.7\%$ of kicks below $50~\kms$ and produces Gaia NSs in the wrong part of the period--eccentricity plane. This advocates for a distinct low-velocity branch of the NS kick distribution.
Importantly, this is {\it not} in tension with isolated-pulsar velocities. 
After accounting for the Gaia NS and Be+NS birth rates, the low-recoil binary-forming channel inferred here needs to contribute only $\sim10\%$ of all NS births (Figure~\ref{fig:parent_kick_law}, right panel),  leaving the full NS kick distribution dominated by the hotter isolated-pulsar population.

\subsection{Physical Interpretation}\label{subsec:physical_interpretation}

\subsubsection{Origin of low kicks}
Across NS populations, there is evidence for both low natal kicks ($0$--$50~\kms$) and high natal kicks ($100$--$1000~\kms$). Which physical channels produce each regime remains an open question.
Large kicks are produced in 3D simulations of standard iron-core collapse when asymmetric ejecta gravitationally accelerate the newborn NS \citep[e.g.,][]{Wongwathanarat2013,Janka2017,Muller2020,Burrows2024,Burrows2024a}.
By contrast, low kicks ($\lesssim50~\kms$) are expected when the explosion is rapid and/or the ejecta mass is small, leaving little time or leverage for large hydrodynamic asymmetries to grow.
This includes electron-capture supernovae (ECSNe) from ONeMg cores, for which multidimensional simulations predict kicks of only a few $\kms$ \citep{Nomoto1984,Nomoto1987,Gessner2018}, ultra-stripped supernovae, whose ejecta masses can be $\lesssim0.1~\msun$ \citep{Tauris2013,Tauris2015}, and possibly some low-mass iron-core collapses in stripped binaries \citep{Podsiadlowski2004,Tauris2017,Janka2017,ODoherty2023}. 
Recent 3D simulations of low-mass progenitors ($\sim9~\msun$) produce total kick speeds of only a few tens of $\kms$ for some rotation rates \citep{Burrows2026}. 

NS masses provide another test of these explosion channels. Recent simulation suites predict that more massive NSs receive larger kicks on average \citep{Burrows2024a,Karim2026}. Gaia NSs, which generally formed through low kicks, should therefore harbor less-massive NSs than populations that received larger kicks, such as HMXBs, LMXBs, or high-$e$ Be+NS binaries.
Existing HMXB measurements may tentatively support this picture. The ten eclipsing HMXBs analyzed by \citet{Falanga2015} have relatively large NS masses (median $1.57~\msun$), and \citet{Fortin2022} infer large natal kicks for HMXBs (median $v_{\rm kick}=116^{+18}_{-15}~\kms$). However, the mass estimates are heavily dependent on models of the donor RVs, donor masses, and eclipse geometry. More work is needed to establish an $M_{\rm NS}$--$v_{\rm kick}$ correlation.

Gaia NS1 makes it difficult to apply such an $M_{\rm NS}$--$v_{\rm kick}$ correlation uniformly. 
It has a mass of $1.88~\msun$, yet its modest eccentricity of $\sim0.1$ \citep[][]{ElBadry2024d} implies it received a low kick. A similar argument disfavors an ECSN origin for the entire Gaia NS population. ECSNe explosions should produce a narrow range of $M_{\rm NS}\approx1.25$--$1.30~\msun$, whereas $12/21$ Gaia NS candidates exceed $1.4~\msun$ \citep{Kitaura2006,Schwab2010,Stevenson2022,ElBadry2024d,ElBadry2026}. Also, the intrinsic Gaia NS mass distribution peaks near $1.3~\msun$ and is broadly similar to those of accreting and recycled NS populations \citep{Shariat2026GaiaNSDemographics}. The low-kick component may therefore include several explosion mechanisms. Larger samples can test whether NS mass and natal kick are correlated.

We also highlight that the inferred Gaia NS natal kicks ($v_{\rm kick}=3$--$30~\kms$) and the systemic velocities they impart ($8$--$30~\kms$) are below the escape speeds of most dense clusters. Thus, such kicks would help retain NSs, consistent with their observations in globular clusters \citep{Katz1975, Drukier1996,Davies1998, Ivanova2008, Pfahl2002a,Ye2019, Katz2021, Kirsten22,Kremer23}. For Gaia NSs specifically, globular cluster models predict formation rates far below the inferred Galactic rate, indicating that cluster dynamics contribute only a small fraction of the observed population \citep[e.g.,][]{Tanikawa2024,SchiebelbeinZwack2026a}.

\subsubsection{Pre-supernova stripping and binary evolution}

The low-kick branches inferred for Be+NS binaries and DNSs are usually interpreted as evidence for strongly stripped progenitors \citep[e.g.,][]{Giacobbo2020,Sgalletta2023}. For example, the low-eccentricity Be+NS systems in \citet{Valli2025} favor pre-supernova masses $M_{\rm He}=2.3^{+1.5}_{-0.9}~\msun$, ejecta masses $\Delta M=0.9^{+1.5}_{-0.9}~\msun$, and ultralow kick velocities ($\lesssim10~\kms$). Their high-eccentricity branch instead favors less-stripped progenitors, $M_{\rm He}=4.9^{+2.4}_{-3.1}~\msun$, together with $\sim100~\kms$ polar kicks. 
Our unified fit allows for both substantially stripped Gaia NS progenitors ($M_{\rm He}=3.08^{+1.34}_{-0.48}~\msun$) and less-stripped progenitors of $M_{\rm He}=4$--$6~\msun$ (see also Appendix \ref{app:high_mhe} for $M_{\rm He}>7~\msun$). The observed Gaia NS kinematics are better reproduced by larger progenitor masses ($M_{\rm He}\sim4$--$6~\msun$; Figure \ref{fig:toomre_ppc}), suggesting that some Gaia NSs may originate from higher mass loss during NS formation.

These masses agree with observational estimates for stripped-envelope (Type Ib/Ic) supernovae and models of binary-stripped helium stars, which generally favor ejecta masses of a few solar masses corresponding to progenitor masses of $\sim2.5$--$5.5~\msun$ \citep[e.g.,][]{Lyman2016,Prentice2019,Woosley2019,Schneider2021,Ertl2020}. The $7$--$10~\msun$ calculation in Appendix~\ref{app:high_mhe} is instead a test of minimal stripping, which could correspond to a hydrogen-rich Type II supernova \citep[e.g.,][]{Valenti2016,Martinez2022}.

Connecting these kick constraints to binary formation histories presents another challenge. Gaia NS progenitors must strip the NS progenitor while leaving a $\sim1~\msun$ companion in a $\sim100$--$1000$ d orbit, {whereas a classical common-envelope phase would likely have led to either merger or a much tighter binary}. One alternative is that the progenitor was only weakly stripped before collapse, or perhaps not at all ($M_{\rm He}\sim7$--$10~\msun$). Appendix~\ref{app:high_mhe} shows that such a high-mass-loss model can match the Gaia period--eccentricity distribution almost as well as the fiducial fit ($\Delta\ln{\cal L}=-0.29$). It is also not unreasonable to assume such wide initial orbits, given the observed population of post-interaction WD+MS binaries in similarly wide orbits \citep[e.g.,][]{Shahaf2024,Yamaguchi2024,Yamaguchi2024b,Yamaguchi2025} and Symbiotic X-ray binaries \citep{Hinkle2006, Yungelson2019, Hinkle2019,Hinkle2024, Nagarajan2024}.

However, the high-$\Delta M$ solution is less efficient than the fiducial low-$\Delta M$ model. 
Because the mass loss is large, most systems are disrupted unless the kick is both modest and strongly retrograde: the surviving systems have $v_{\rm kick}=51~\kms$ ($36$--$77~\kms$) and $\theta_v=157^\circ$ ($147^\circ$--$168^\circ$). 
Thus, the Gaia data do not rule out a weakly stripped RSG- or super-AGB progenitor, but this channel likely requires the progenitor to avoid engulfing the au-scale companion before collapse. 
If such progenitors expand to several hundred $R_\odot$ and cause significant orbital shrinkage or mergers, the channel becomes strongly disfavored. 
Longer-period Gaia NS binaries in {\it Gaia} DR4 will provide a direct test of this scenario.

\subsection{Rocket kicks}\label{subsec:rocket_kicks}

\citet{Hirai2024} proposed that a delayed rocket-like acceleration could reduce the eccentricities of Gaia NSs formed from compact post-common-envelope binaries ($P_{\rm orb}\approx1$ d).  Their simulations produce wide, low-eccentricity systems with an efficiency $\eta_{\rm roc}\approx5\times10^{-4}$--$10^{-3}$. Compared with the fraction of NSs that are in Gaia NS binaries, $q_{\it Gaia}\approx5\times10^{-4}$ \citep{Shariat2026GaiaNSDemographics}, this requires
\begin{equation}
    f_{\rm ch}\approx\frac{q_{\it Gaia}}{\eta_{\rm roc}}
    \approx0.47\text{--}0.94.
\end{equation}
Thus, if rockets produced the full Gaia NS population, $47$--$94\%$ of all NS births would first need to pass through this specific compact binary configuration.  This high required channel fraction disfavors rockets as the dominant origin of Gaia NSs.

\citet{Baibhav2026} explored a broader range of progenitors and found that rockets can improve models starting from compact pre-supernova binaries, whereas wider progenitors ($P\sim100$~days) frequently reproduce the Gaia NSs without a rocket kick needed.  Their preferred pre-supernova separations of approximately $0.5$--$3$ au are consistent with our results.  Moreover, they note that producing rocket velocities $\gtrsim30~\kms$ as argued by \citet{Hirai2024} requires unusually rapid NS birth spins \citep[e.g.,][for observed birth spins]{Lai2001,Janka2022} and restrictive magnetospheric conditions \citep{Baibhav2026}.  In summary, rocket kicks are not required to explain either the orbits or inferred birth rate of Gaia NSs, especially if the progenitors were already in wide orbits (Section~\ref{subsubsec:comp_to_GaiaNS}).

\subsection{Living progenitors}\label{subsec:stripped_star_progenitors}

Our kick model directly predicts the progenitors of Gaia NSs. 
In the isolated-binary picture considered here, the immediate pre-supernova binaries contain stripped helium stars with characteristic mass $M_{\rm He}\sim2.3-5~\msun$ and $\sim1~\msun$ main-sequence companions in roughly au-scale orbits. These {progenitor masses} lie in the intermediate-mass stripped star regime, $M_{\rm strip}\sim1$--$8~\msun$, between ordinary hot subdwarfs and classical Wolf-Rayet stars \citep{Gotberg2018,Drout2023Stripped,Gotberg2023Stripped,HovisAfflerbach2025}. Such stars are predicted and observed to be hot, UV-bright, helium-rich, and hydrogen-poor, with typical $T_{\rm eff}\sim50$--$100~{\rm kK}$, $\log g\sim5$, and radii of order $0.5$--$1~\rsun$ \citep[e.g.,][]{Gotberg2018,Drout2023Stripped,Gotberg2023Stripped}. A young stripped star should outshine a $\sim1~\msun$ companion in the UV, making such systems easier to identify than analogous binaries with luminous B-type companions, where the stripped star signature is more strongly diluted \citep[e.g.,][]{Drout2023Stripped,Blomberg2025Stripped,Ludwig2025SUMS}.

The most direct search for Gaia NS progenitors is therefore for hot $M_{\rm He}\gtrsim2~\msun$ helium stars with low-mass companions and periods of $\sim50$ to $\sim10^3$ days. 
Recent RV monitoring has revealed intermediate-mass stripped stars
with optically faint companions and orbital periods extending to
$\sim300$ days \citep{Laroche2026}. These companions could be low-mass stars, consistent with the progenitors of Gaia NSs.
A non-detection in targeted UV-selected spectroscopic searches would suggest that the progenitors are shorter-lived, cooler \citep[e.g.,][]{DuttaKlencki2024}, or not well described by the simple isolated He+MS channel considered here.

The observed Gaia NS birth rate \citep[$\mathcal{R}_{\rm Gaia~NS}\sim10^{-5}~{\rm yr^{-1}}$;][]{Shariat2026GaiaNSDemographics} sets the required `supply' of pre-supernova progenitors. If a fraction $\eta_{\it Gaia}=0.04$ of He+MS systems in the relevant pre-supernova orbits survive the kick and land in Gaia NS-like orbits, then the required progenitor birth rate is $\mathcal{R}_{\rm He+MS}\approx\mathcal{R}_{\rm Gaia~NS}/\eta_{\it Gaia}\approx2\times10^{-4}~{\rm yr^{-1}}$. 

\begin{figure}
    \centering
    \includegraphics[width=0.99\columnwidth]{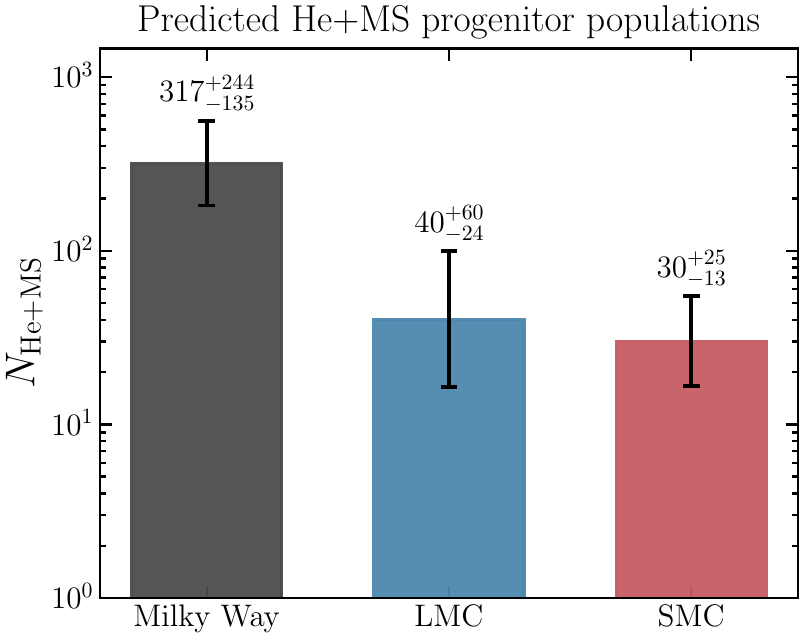}
    \caption{Predicted population of intermediate-mass stripped star binaries, the progenitors of Gaia NSs.
    Bars show the posterior distributions of the number of He+MS binaries presently in the au-scale pre-supernova phase, based on our best-fit kick model.  We show results for the Milky Way (black), LMC (blue), and SMC (red).}
    \label{fig:gaia_ns_progenitor_counts}
\end{figure}

To convert this birth rate into a present-day population, we integrate the inferred He-star mass distribution over the mass-dependent stripped star lifetimes calculated by \citet{HovisAfflerbach2025} at $Z=0.014$, $0.006$, and $0.002$ for the Milky Way, LMC, and SMC, respectively.  We scale the LMC progenitor rate by its star formation rate relative to the Milky Way \citep{HarrisZaritsky2009,LicquiaNewman2015}.  For the SMC, because the star formation rate varies strongly over the past $\sim100$~Myr, we instead use the resolved history from \citet{Harris2004}. For each $M_{\rm He}$, we evaluate the SMC star formation rate at the corresponding lookback time, $t_{\rm MS}+t_{\rm strip}/2$, and integrate over the parent mass distribution.  
These calculations predict $317^{+244}_{-135}$ He+MS progenitors in the Milky Way, $40^{+60}_{-24}$ in the LMC, and $30^{+25}_{-13}$ in the SMC at present (Figure~\ref{fig:gaia_ns_progenitor_counts}).
This calculation accounts for metallicity through the stripped star lifetimes, but assumes that the efficiency of forming such He+MS systems per unit stellar mass is metallicity invariant.

The Gaia NS birth rate can also be recast to constrain the number of massive star binaries with initial mass ratios $q<0.1$. In Appendix~\ref{app:lowq_progenitor}, we show that the observed Gaia NS abundance requires a $\sim7\%$ supply of massive-star binaries with $q<0.1$ in au-scale separations to achieve the abundance of Gaia NSs in our Galaxy.

\section{Summary and Conclusions}\label{sec:conclusions}

{\it Gaia} DR3 revealed a new class of neutron stars in au-scale binaries with solar-type companions \citep[][]{ElBadry2024b,ElBadry2024d,ElBadry2026}.  
Their origin remains puzzling: the orbits are close enough that the NS progenitor likely interacted with the present companion, but too wide to be an obvious outcome of common-envelope evolution and too fragile to survive large natal kicks.  
Yet this detached orbital architecture with low-mass companions makes them uniquely sensitive probes of NS natal kicks.  
Once the NS formed, the orbit has remained largely the same, so the present-day periods and eccentricities provide pristine records of the supernova mass loss and natal kick that occurred at NS formation.

Using the intrinsic period--eccentricity distribution and space density of Gaia NSs \citep{Shariat2026GaiaNSDemographics}, we infer the kicks and progenitors required to form them.   We then jointly model Gaia NSs, Be+NS binaries, and isolated pulsars to construct a concordance model of NS natal kicks. Our main conclusions are summarized as follows:

\begin{enumerate}
    \item \textit{Gaia NS kicks:} The Gaia NS population favors a broad parent kick distribution with median speed
    $v_{\rm kick}=24.2^{+32.6}_{-17.9}~\kms$.  Systems that {\it survive} as Gaia NSs
    had progenitor masses $M_{\rm He}=3$--$6.5~\msun$ and kicks of
    $v_{\rm kick}=29^{+14}_{-20}~\kms$, preferentially anti-aligned with the pre-supernova orbital motion
    (see, Figure~\ref{fig:parent_fourpanel}; Appendix~\ref{app:survivor_distributions}).

    \item \textit{A universal kick law:} 
    The orbits and birth rates of Gaia NS and Be+NS binaries, together with the velocities of young isolated pulsars, are consistent with a single bimodal kick distribution for all NS births.
    This distribution contains a low-kick component, with $\mu_{\ln v}=1.87^{+0.65}_{-0.53}$ and $\sigma_{\ln v}=0.55^{+0.85}_{-0.44}$, and the high-kick component inferred from young pulsars (Figure~\ref{fig:parent_fourpanel}).
    The low-kick component accounts for only $f_{\rm low}=12.6^{+8.2}_{-4.4}\%$ of all NS births (Figure~\ref{fig:parent_kick_law}), but is overrepresented among bound NS binaries.
    It therefore produces $\sim90\%$ of Gaia NS binaries and $\sim50\%$ of Be+NS binaries (Figure~\ref{fig:joint_pe_check}).
    Thus, a relatively small low-kick component can explain a large fraction of the surviving binary populations while remaining consistent with isolated-pulsar constraints (Figure~\ref{fig:previous_kick_constraints}).

    \item \textit{Kinematics:} 
    \begin{enumerate} 
    
    \item The space velocities of Gaia NS candidates are generally similar to, or only mildly elevated above, their neighboring stars (as depicted in Figure~\ref{fig:toomre}).
    Of the $24$ non-halo binaries, $12$ have no measurable excess above the local velocity dispersion, while the $12$ systems {\it with} an excess span $v_{\rm kick,min}=10$--$41~\kms$. The kinematics are consistent with predictions from the unified kick model if for NS progenitors with $M_{\rm He}\sim4$--$6~\msun$ (Figure~\ref{fig:toomre_ppc}).
            
    \item NS-LMXBs are moving $2$--$3\times$ faster than Gaia NSs on average (Figure~\ref{fig:lmxb_velocity_compare}), which can be explained by the same $v_{\rm kick}$ distribution if NS-LMXBs come from tighter pre-supernova binaries (e.g., Equation (\ref{eq:systemic_recoil})).

    \item High-$e$ Be+NS binaries have a median peculiar velocity of $24~\kms$, compared with $11~\kms$ for the low-$e$ branch, consistent with the model prediction that the high-$e$ systems are typically $\sim1.5$--$3\times$ faster (Figure~\ref{fig:bens_kinematics}).  

    \end{enumerate}

    \item \textit{NSs in triples:} We consider the survival of tertiary companions to Gaia NSs under the inferred kicks.  
    The fractions of Gaia NSs predicted to retain
    stable tertiaries with $a_{2,\rm final}>10$, $50$, $500$, and $1000$ au are $6.6\%$, $5.5\%$, $1.5\%$, and $0.86\%$, respectively (Figure~\ref{fig:tertiary_survival}).  Tertiaries that survive the kicks are preferentially massive stars in close initial outer orbits, typically with $M_3\gtrsim3~\msun$ and $a_{2,i}\lesssim500$~au (Table~\ref{tab:tertiary_summary}).

    Be+NS binaries retain wide tertiaries more often than Gaia NS binaries because their more massive inner binaries are harder to unbind (Figure~\ref{fig:bens_tertiary_survival}). We find that $\sim15\%$ retain a stable companion beyond $50$~au. Under the assumed mass ratio distribution, these tertiaries typically have $M_3\sim10~\msun$ (Table~\ref{tab:tertiary_summary}).

    \item \textit{Gaia NS Progenitors:} In the isolated-binary channel, Gaia NSs descend from hot intermediate-mass stripped stars ($M_{\rm He}\sim2$--$8~\msun$) with $\sim1~\msun$ companions in $\sim50$--$1000$~day orbits. 
    The best-fit kick model predicts $200$--$500$ such He+MS binaries in the Milky Way, $20$--$100$ in the LMC, and $20$--$50$ in the SMC, assuming that their formation is weakly dependent on metallicity (Figure~\ref{fig:gaia_ns_progenitor_counts}).
    The observed abundance of Gaia NSs also requires that $\sim1$--$10\%$ of all OB stars have low-mass companions ($q<0.1$) in au-scale orbits (Appendix~\ref{app:lowq_progenitor}).

\end{enumerate} 

Overall, our results suggest that the orbits, kinematics, and birth rates of NS populations can be reconciled with a shared bimodal natal kick law once binary survival and observational selection effects are accounted for. Larger Gaia NS and Be+NS samples with precise orbits and systemic velocities will test and refine this picture, while searches for stripped star progenitors and tertiary companions will constrain their formation histories. 
Interpreting these empirical constraints through models of binary evolution and core collapse will help connect the inferred kick distribution to the physics of neutron star formation {\citep[e.g.,][]{Podsiadlowski2004,Giacobbo2020,TaurisvandenHeuvel2023,Janka2017,Burrows2024a}}.

\section*{Acknowledgments} \label{acknowledgments}
{We thank the anonymous referee for helpful comments that improved the clarity and presentation of the manuscript.}
C.S. acknowledges support from the Department of Energy Computational Science Graduate Fellowship.
This material is based upon work supported by the U.S. Department of Energy, Office of Science, Office of Advanced Scientific Computing Research, under Award Number DE-SC0026073. 
This research was supported in part by grant NSF PHY-2309135 to the Kavli Institute for Theoretical Physics (KITP) and by NSF grant AST-2540180.
S.N. thanks Howard and Astrid Preston for their generous support.

This work presents results from the European Space Agency (ESA) space mission {\it Gaia}. {\it Gaia} data are being processed by the {\it Gaia} Data Processing and Analysis Consortium (DPAC). Funding for the DPAC is provided by national institutions, in particular the institutions participating in the {\it Gaia} MultiLateral Agreement (MLA). The {\it Gaia} mission website is \url{https://www.cosmos.esa.int/Gaia}. The {\it Gaia} archive website is \url{https://archives.esac.esa.int/Gaia}.

\software{This work made use of \texttt{OverCite} \citep{Shariat2026b}, an in-editor citation tool for \LaTeX. This work made use of the following software packages: \texttt{Jupyter} \citep{2007CSE.....9c..21P,kluyver2016jupyter}, \texttt{matplotlib} \citep{Hunter:2007}, \texttt{numpy} \citep{numpy}, \texttt{pandas} \citep{mckinney-proc-scipy-2010,pandas_17806077}, \texttt{python} \citep{python}, \texttt{scipy} \citep{2020SciPy-NMeth,scipy_17467817}, \texttt{Cython} \citep{cython:2011}, \texttt{emcee} \citep{ForemanMackey2013,emcee_10996751}, \texttt{h5py} \citep{collette_python_hdf5_2014,h5py_7560547}, \texttt{schwimmbad} \citep{schwimmbad}, \texttt{seaborn} \citep{Waskom2021},  and \texttt{tqdm} \citep{tqdm_14231923}.
Software citation information aggregated using \texttt{\href{https://www.tomwagg.com/software-citation-station/}{The Software Citation Station}} \citep{software-citation-station-paper,software-citation-station-zenodo}.
}

\appendix
\twocolumngrid
\makeatletter
\@addtoreset{equation}{section}
\renewcommand{\theequation}{\thesection\arabic{equation}}
\renewcommand{\theHequation}{\thesection.\arabic{equation}}
\makeatother

\section{Instantaneous supernova calculation}\label{app:sn_formalism}
\subsection{Binaries}%

{We use the instantaneous-supernova formalism \citep{Blaauw1961,Hills1983,Kalogera1996,Kalogera1998,TaurisvandenHeuvel2023}. The mass loss and natal kick are assumed to occur on a timescale short compared with the orbital period. Stellar positions and the companion velocity remain unchanged during the explosion. We treat the mass loss as symmetric in the progenitor's rest frame, with explosion asymmetries represented by the natal kick, and neglect ejecta--companion interaction.

Let $\mathbf r=\mathbf r_{\rm He}-\mathbf r_2$ and $\mathbf v=\mathbf v_{\rm He}-\mathbf v_2$ be the relative position and velocity just before the explosion, with $r=|\mathbf r|$. The total binary mass changes from $M_{\rm pre}=M_{\rm He}+M_2$ to $M_{\rm post}=M_{\rm NS}+M_2$, with $\Delta M=M_{\rm pre}-M_{\rm post}$. The new relative state is
\begin{equation}
    \mathbf r_{\rm post}=\mathbf r,\qquad
    \mathbf v_{\rm post}=\mathbf v+\mathbf v_{\rm kick}.
    \label{eq:sn_relative_velocity}
\end{equation}
The orbital energy and angular momentum per unit reduced mass are
\begin{equation}
\begin{aligned}
    \varepsilon_{\rm post}&=\frac{|\mathbf v_{\rm post}|^2}{2}
                         -\frac{GM_{\rm post}}{r},\\
    \mathbf h_{\rm post}&=\mathbf r\times\mathbf v_{\rm post}.
\end{aligned}
\label{eq:sn_energy_angular_momentum}
\end{equation}
For bound, nonradial solutions ($\varepsilon_{\rm post}<0$ and $e_{\rm post}<1$), the new orbital elements are
\begin{equation}
\begin{aligned}
    a_{\rm post}&=-\frac{GM_{\rm post}}{2\varepsilon_{\rm post}},\\
    \mathbf e_{\rm post}&=\frac{\mathbf v_{\rm post}\times\mathbf h_{\rm post}}
                              {GM_{\rm post}}-\frac{\mathbf r}{r},\\
    e_{\rm post}&=|\mathbf e_{\rm post}|,\qquad
    P_{\rm post}=2\pi\sqrt{\frac{a_{\rm post}^3}{GM_{\rm post}}}.
\end{aligned}
\label{eq:sn_orbital_elements}
\end{equation}
For circular progenitors, $r=a_{\rm pre}=[GM_{\rm pre}(P_{\rm preSN}/2\pi)^2]^{1/3}$, $|\mathbf v|=\sqrt{GM_{\rm pre}/a_{\rm pre}}$, and $\mathbf r\cdot\mathbf v=0$. In this case, setting the natal kick to zero recovers $e_{\rm post}=\Delta M/M_{\rm post}$ for bound systems. The calculation above also applies to eccentric progenitors, using their position and velocity at the time of the explosion.

The systemic recoil is given by Equation~\ref{eq:systemic_recoil}; in the pre-supernova center-of-mass frame, $\mathbf v_{\rm He}=(M_2/M_{\rm pre})\mathbf v$. We then apply the period, eccentricity, and observational selections described in Section~\ref{sec:methods}.
}

\subsection{Hierarchical triples}%

{For hierarchical triples, the tertiary orbits the inner binary's center of mass in the 'outer orbit.' We therefore treat the inner binary as a point mass at its center of mass when calculating the outer orbit's response to supernova mass loss and recoil \citep[see][for a full list of equations]{Lu2019}.
Let $\mathbf R$ and $\mathbf V$ be the tertiary's position and velocity relative to the inner binary's pre-supernova center of mass. The tertiary receives no direct impulse. Although the stellar positions remain fixed, mass loss shifts the inner center of mass by $-\Delta M M_2\mathbf r/(M_{\rm pre}M_{\rm post})$. The new outer relative state is therefore
\begin{equation}
\begin{aligned}
    \mathbf R_{\rm post}&=\mathbf R+
      \frac{\Delta M M_2}{M_{\rm pre}M_{\rm post}}\mathbf r,\\
    \mathbf V_{\rm post}&=\mathbf V-\mathbf V_{\rm sys}.
\end{aligned}
\label{eq:sn_outer_state}
\end{equation}
The outer orbital elements follow from Equations~\ref{eq:sn_energy_angular_momentum}--\ref{eq:sn_orbital_elements}, using $(\mathbf R_{\rm post},\mathbf V_{\rm post})$ and total mass $M_{\rm post}+M_3$, where $M_3$ is the tertiary mass. The post-supernova mutual inclination is
\begin{equation}
\begin{aligned}
    i_{\rm mut,post}&=\arccos\!\left(
    \frac{\mathbf h_{\rm post}\cdot\mathbf h_{2,\rm post}}
    {|\mathbf h_{\rm post}|\,|\mathbf h_{2,\rm post}|}\right),\\
    \mathbf h_{2,\rm post}&=\mathbf R_{\rm post}\times\mathbf V_{\rm post}.
\end{aligned}
\label{eq:sn_mutual_inclination}
\end{equation}
We require both orbits to remain bound and apply the hierarchy cut \citep{Naoz2013}
\begin{equation}
    \frac{a_1}{a_2}\frac{e_2}{1-e_2^2}<0.1
    \label{eq:sn_hierarchy_cut}
\end{equation}
and the \citet{Mardling2001} stability criterion,
\begin{equation}
\begin{split}
    \frac{a_2}{a_1}>
    &\,2.8\left(1+\frac{M_3}{M_{\rm in}}\right)^{2/5}\\
    &\times\frac{(1+e_2)^{2/5}}{(1-e_2)^{6/5}}
    \left(1-\frac{0.3i_{\rm mut}}{\pi}\right).
\end{split}
\label{eq:sn_stability_cut}
\end{equation}
Here $a_1$ and $a_2$ are the inner and outer semimajor axes, $e_2$ is the outer eccentricity, and $i_{\rm mut}$ is in radians. We apply these cuts before and after the supernova, with $M_{\rm in}=M_{\rm pre}$ and $M_{\rm post}$, respectively. We do not follow subsequent three-body or stellar evolution.
}

\section{Survivor-weighted kick distributions}\label{app:survivor_distributions}

\begin{figure}
    \centering
    \includegraphics[width=0.99\linewidth]{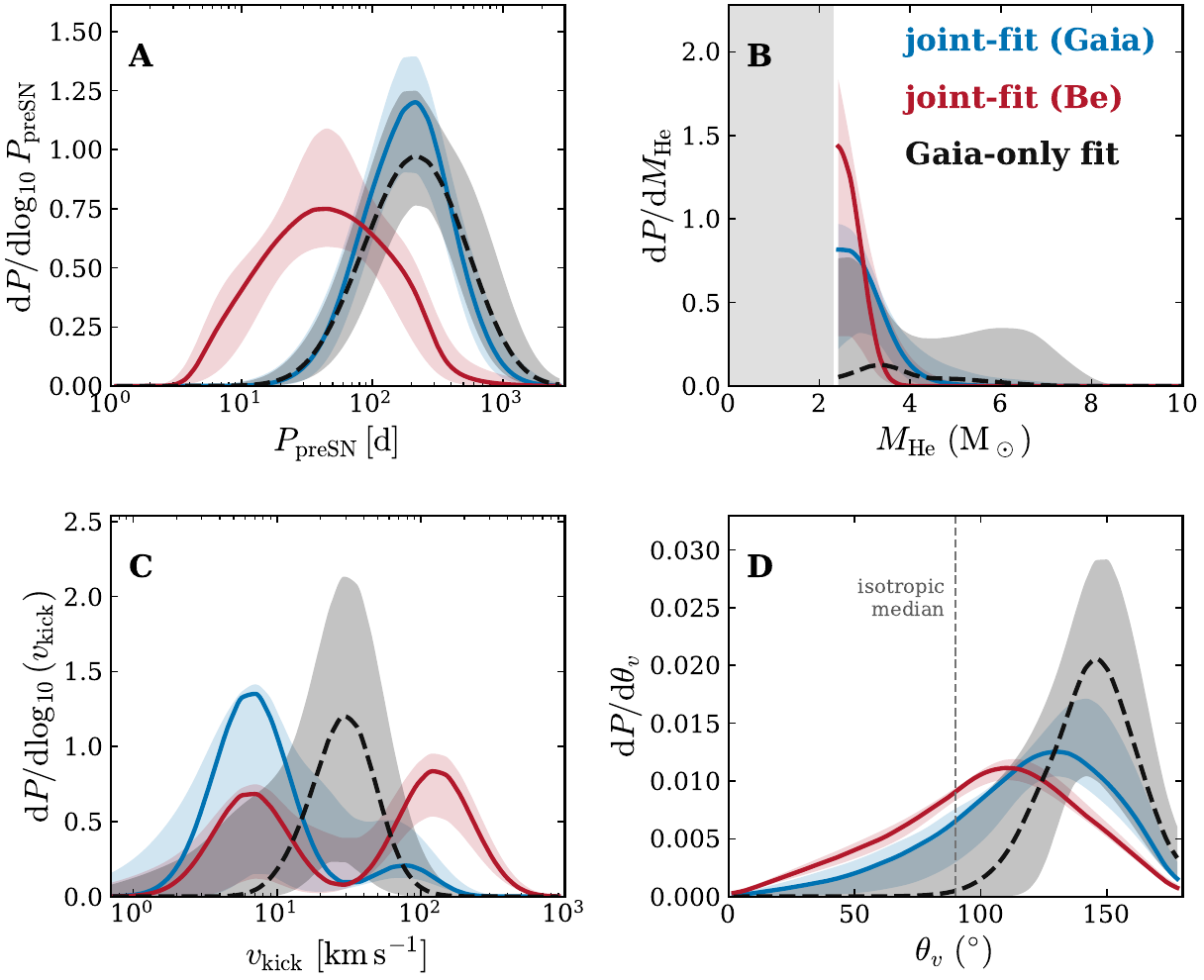}
    \caption{
    Survivor-weighted distributions of progenitors and natal kicks. Panels A--D show the pre-supernova periods, {progenitor masses}, kick speeds, and kick directions among systems that remain bound and enter
    the Gaia NS or Be+NS samples.  Blue and red show the Gaia NS and Be+NS populations from the unified fit, while dashed black shows the Gaia NS-only fit.  The Gaia curves include the astrometric selection function, while the Be+NS curves assume a complete sample. Shaded regions show the 16th--84th posterior percentiles.  The parent kick directions are isotropic, so the retrograde preference among Gaia NS survivors is due to survivor bias.  Comparison with Figure~\ref{fig:parent_fourpanel} shows how the two binary populations select different subsets of the same parent distributions.
    }
    \label{fig:survivor_fourpanel_appendix}
\end{figure}

Figure~\ref{fig:survivor_fourpanel_appendix} shows the survivor-weighted distributions from the Gaia NS-only (black) and unified (colored) fits after selecting binaries that remain bound and enter the relevant observable sample. The corresponding parent progenitor distributions are shown in Figure~\ref{fig:parent_fourpanel}. The selected Gaia NS systems favor pre-supernova periods of a few hundred days and {progenitor masses} centered near $3~\msun$. Most arise from the low-kick component, giving a median kick speed of $6.8^{+12.0}_{-2.6}~\kms$. Their kick directions are preferentially opposite the pre-supernova orbital motion. Because the kick directions are isotropic in the parent population, this retrograde preference arises from survivors preferentially selecting anti-aligned kicks.

The Be+NS survivors select shorter pre-supernova periods, peaking near $50$~d, and {progenitor masses} concentrated near $2.6~\msun$. Both kick components contribute substantially, producing a bimodal selected distribution with a median of $47.9^{+27.5}_{-29.4}~\kms$. The kick directions are broader than for Gaia NSs and extend across the isotropic median of $90^\circ$. The more massive Be companion allows binaries to survive over a wider range of kick directions. Figure~\ref{fig:survivor_fourpanel_appendix} illustrates that the Gaia NS and Be+NS populations select different subsets of the same parent kick law.

\section{Effect of pre-supernova eccentricity}\label{app:presn_eccentricity}

The fiducial calculation assumes circular pre-supernova orbits, motivated by tidal interactions with the NS progenitor.  We test this assumption by repeating the unified Gaia--Be--pulsar fit (Section \ref{subsec:joint_method}) with $e_{\rm preSN}\sim\mathcal{U}(0,1)$.  For each supernova explosion, we draw the mean anomaly uniformly over $0$--$2\pi$ and solve Kepler's equation, corresponding to a supernova at a random time in the orbit. We then apply the instantaneous kick equations to calculate the post-supernova orbit. The progenitor priors, kick-law parameterization, selection functions, sample definitions, and likelihood terms are the same as the fiducial fit.

\begin{figure}
    \centering
    \includegraphics[width=0.99\linewidth]{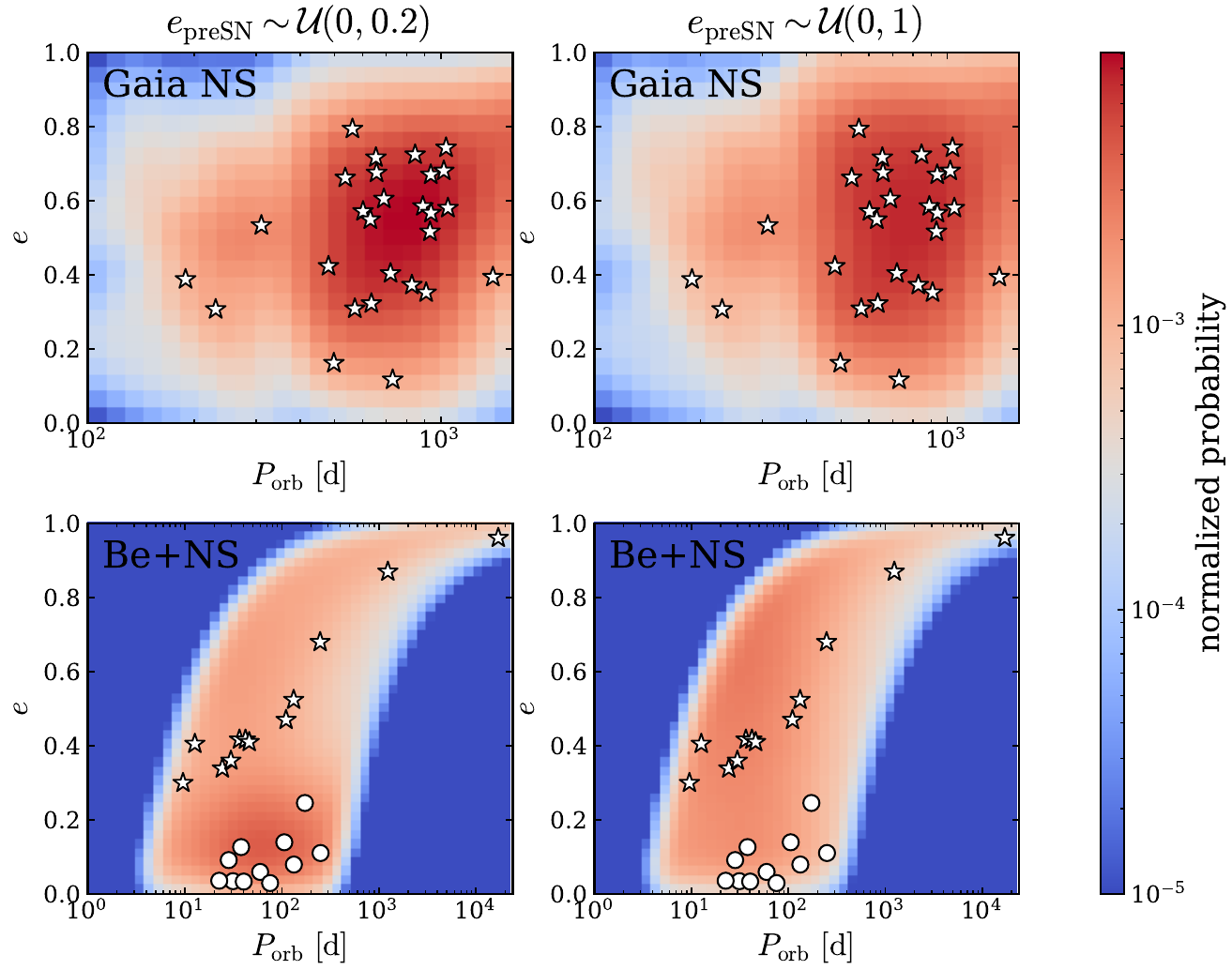}
    \caption{Predicted period--eccentricity distributions with pre-supernova eccentricities drawn from $e_{\rm preSN}\sim\mathcal{U}(0,0.2)$ and $e_{\rm preSN}\sim\mathcal{U}(0,1)$. Formatting is the same as Figure \ref{fig:presn_eccentricity_ppc}.  Each panel is normalized to unit probability over its displayed grid.  Both models reproduce the broad Gaia NS distribution, but the $\mathcal{U}(0,1)$ model provides a poorer description of the Be+NS period--eccentricity sequence.}
    \label{fig:presn_eccentricity_ppc}
\end{figure}

Allowing eccentric progenitors changes the inferred low-kick component.  Its
median speed shifts from $27.1^{+26.4}_{-15.9}~\kms$ for circular progenitors to $4.1^{+6.8}_{-2.6}~\kms$ for $e_{\rm preSN}\sim\mathcal{U}(0,1)$. The Gaia NS formation efficiency increases from $0.037$ to $0.102$, while the Be+NS efficiency decreases from $0.577$ to $0.454$.  However, the low-kick fraction among all NS births remains nearly unchanged: $f_{\rm low,all}=0.0094^{+0.0143}_{-0.0048}$ for circular progenitors and $0.0072^{+0.0024}_{-0.0021}$ for the flat eccentricity model.  The fit is marginally worse than the circular model. The median Gaia NS log-likelihood changes by only $-0.64$, but the Be+NS term decreases by $-6.59$.  

The uniform-$e$ prior also includes many progenitors that would interact with the NS progenitor.  Neither this experiment nor the circular fiducial model applies a pre-supernova periastron cut in either mode.  Under the flat-eccentricity fit, $48\%$ of Gaia NS and $75\%$ of Be+NS parent progenitors have $a_{\rm preSN}(1-e_{\rm preSN})<100~\rsun$; among the selected survivors, the corresponding fractions are $89\%$ and $64\%$. The $\mathcal{U}(0,1)$ calculation is mainly a stress test that does not assume a physical binary formation model.  A physical fit for the pre-supernova eccentricity will require the progenitor radii and prior binary interaction to be modeled simultaneously.

As another comparison, we repeat the calculation with $e_{\rm preSN}\sim\mathcal{U}(0,0.2)$, similar to the eccentricities observed among stripped star binaries \citep{Valli2025}.  This model gives a
low-component median speed of $25.1^{+15.4}_{-14.7}~\kms$, consistent with the circular result (Figure \ref{fig:parent_fourpanel}), and $f_{\rm low,all}=0.0089^{+0.0096}_{-0.0039}$.  Its posterior-predictive $P$--$e$ distributions are effectively the same as the circular model. Our results are therefore unchanged for moderately eccentric pre-supernova binaries $e\sim\mathcal{U}(0,0.2)$.

\section{Kinematic Comparison}\label{app:toomre}

\subsection{Gaia NS}\label{subapp:toomre_gaians}

\vspace{-0.8cm}
\begin{deluxetable*}{lrrrrrrrr}
\tablecaption{Kinematics of the 27 Gaia NS candidates compared to their local stellar population. $V_{\rm pec,local}$ is measured relative to the median azimuthal velocity of the local Gaia DR3 control sample; $V_{68}$, $V_{90}$, and $V_{95}$ are percentiles of that local-control distribution. We define the present-day local velocity excess as $V_{\rm exc,68}=\max(0,V_{\rm pec,local}-V_{68})$. The primed columns use the median-centered perpendicular velocity statistic defined below.\label{tab:gaia_ns_kinematics}}
\tablehead{
\colhead{System} & \colhead{$V_{\rm pec,local}$} & \colhead{$V_{68}$} & \colhead{$V_{90}$} & \colhead{$V_{95}$} & \colhead{$v_{\rm kick,min}$} & \colhead{Percentile} & \colhead{$v^{\prime}_{\rm kick,min}$} & \colhead{Percentile$^{\prime}$} \\
\colhead{} & \colhead{(km s$^{-1}$)} & \colhead{(km s$^{-1}$)} & \colhead{(km s$^{-1}$)} & \colhead{(km s$^{-1}$)} & \colhead{(km s$^{-1}$)} & \colhead{(\%)} & \colhead{(km s$^{-1}$)} & \colhead{(\%)}
}
\startdata
J0553-1349 & 20.1 & 45.3 & 73.1 & 84.6 & 0.0 & 22.0 & 0.0 & 5.2 \\
J0230+5950 & 27.2 & 48.2 & 70.0 & 79.6 & 0.0 & 33.6 & 0.0 & 0.0 \\
J1150-2203 & 28.8 & 51.2 & 79.0 & 96.6 & 0.0 & 31.7 & 0.6 & 68.9 \\
J2244-2236 & 34.6 & 55.7 & 83.6 & 101.8 & 0.0 & 35.4 & 0.0 & 0.5 \\
Gaia DR3 809741149368202752 & 35.2 & 55.0 & 81.6 & 96.7 & 0.0 & 35.2 & 0.0 & 0.4 \\
J2057-4742 & 36.2 & 51.8 & 78.3 & 95.9 & 0.0 & 41.4 & 0.0 & 39.1 \\
J1553-6846 & 38.2 & 49.1 & 77.1 & 90.6 & 0.0 & 50.9 & 0.0 & 58.1 \\
J2145+2837 & 39.1 & 46.6 & 76.9 & 90.4 & 0.0 & 56.0 & 5.7 & 75.3 \\
Gaia DR3 1695294922548180224 & 40.9 & 55.6 & 82.8 & 102.2 & 0.0 & 46.6 & 0.0 & 11.8 \\
J2102+3703 & 42.7 & 48.1 & 73.8 & 86.0 & 0.0 & 61.5 & 7.0 & 78.1 \\
Gaia DR3 4240540718818313984 & 46.5 & 49.6 & 75.7 & 89.2 & 0.0 & 63.8 & 0.0 & 49.4 \\
J0036-0932 & 49.9 & 56.6 & 83.8 & 97.2 & 0.0 & 58.8 & 0.0 & 30.0 \\
J0639-3655 & 58.5 & 48.7 & 71.5 & 84.3 & 9.8 & 81.1 & 3.4 & 72.9 \\
Gaia DR3 1581117310088807552 & 60.0 & 47.0 & 72.3 & 86.4 & 13.0 & 80.8 & 27.1 & 92.9 \\
J0217-7541 & 65.0 & 54.0 & 80.4 & 94.4 & 11.0 & 79.0 & 4.2 & 74.8 \\
J0634+6256 & 66.1 & 48.5 & 72.0 & 83.0 & 17.6 & 86.2 & 4.3 & 76.8 \\
J0824+5254 & 73.4 & 49.2 & 75.0 & 91.8 & 24.2 & 89.3 & 16.2 & 85.4 \\
J0003-5604 & 77.0 & 53.7 & 85.2 & 102.0 & 23.4 & 86.8 & 37.7 & 94.3 \\
J1733+5808 & 80.8 & 50.9 & 75.9 & 90.0 & 29.9 & 91.3 & 18.1 & 88.8 \\
Gaia DR3 747174436620510976 & 82.0 & 49.8 & 78.8 & 91.4 & 32.2 & 91.2 & 19.3 & 88.1 \\
Gaia DR3 6054379247042197504 & 82.8 & 49.2 & 71.6 & 86.4 & 33.6 & 94.3 & 20.8 & 91.1 \\
J1048+6547 & 85.5 & 57.4 & 85.0 & 104.7 & 28.1 & 90.3 & 12.8 & 83.5 \\
J0742-4749 & 86.0 & 46.7 & 69.3 & 82.6 & 39.3 & 95.8 & 27.6 & 94.5 \\
Gaia DR3 1144019690966028928 & 90.4 & 49.8 & 75.1 & 85.4 & 40.6 & 95.7 & 35.0 & 95.2 \\
J0152-2049$^{\ast}$ & 282.8 & 53.4 & 79.5 & 92.9 & 229.3 & 99.9 & 228.2 & 99.9 \\
J1432-1021$^{\ast}$ & 336.9 & 58.8 & 86.0 & 100.9 & 278.1 & 99.9 & 273.0 & 99.9 \\
J1739+4502$^{\ast}$ & 340.7 & 53.7 & 81.0 & 96.9 & 287.0 & 100.0 & 276.4 & 99.9 \\
\enddata
\tablecomments{*halo orbits}

\end{deluxetable*}

For the $27$ Gaia NS candidates, we construct the local control samples used in Section~\ref{subsec:toomre_main}, following \citet{Nagarajan2025a}.
{The coordinates, proper motions, systemic velocities, and uncertainties for each Gaia NS candidate are taken from the discovery catalog \citep[][]{ElBadry2024b,ElBadry2026}. 
We query Gaia DR3 stars with measured radial velocities in {\tt gaiadr3.gaia\_source\_lite} within an angular radius
\begin{equation}
    \theta_{500}=\tan^{-1}\left(\frac{500~{\rm pc}}{D}\right),
\end{equation}
where $D$ is the target distance, selecting a projected radius of 500 pc. We use the Bailer-Jones geometric distances \citep{BailerJones2021}, with initial window $(D-\Delta D_-,D+\Delta D_+)$ set by the distance posterior. The query uses the corresponding parallax interval, expanded by $15\%$ on each side to avoid excluding valid control stars.}

The local-control stars must have finite radial velocities, ${\tt parallax\_over\_error}\ge3$, and radial velocity uncertainty below $5~\kms$.  After fetching the {\tt parallax\_pmra\_corr}, {\tt parallax\_pmdec\_corr}, and {\tt pmra\_pmdec\_corr} covariance terms from {\tt gaiadr3.gaia\_source}, we propagate the parallax--proper-motion covariance into the tangential-velocity uncertainty and require $\sigma_{v_{\rm tan}}<30~\kms$. In the final Gaia NS run, all 27 systems have at least $200$ control stars.

We compute Galactocentric cylindrical velocities for the target and for every local control star using \texttt{astropy}.  The adopted Galactocentric frame has $R_0=8.122~{\rm kpc}$, $z_\odot=20.8~{\rm pc}$, and solar cylindrical velocity $(V_R,V_\phi,V_Z)=(-12.9,245.6,7.78)~\kms$ \citep[e.g.,][]{Bovy2016}.  For each Gaia NS candidate, we draw $10{,}000$ Monte Carlo realizations of the distance, proper motion, and systemic velocity. The velocity and error bars in Figure~\ref{fig:toomre} give the median and 16th--84th percentile range of these Monte Carlo velocities.

Subtract the local median azimuthal motion ($\widetilde{V}_{\phi,{\rm local}}$), we define
\begin{equation}
    V_{\rm pec,local}
    =
    \left[
    V_R^2+V_Z^2+
    \left(V_\phi-\widetilde{V}_{\phi,{\rm local}}\right)^2
    \right]^{1/2}.
\end{equation} 
Here, $V_{68}$, $V_{90}$, and $V_{95}$ are the 68th, 90th, and 95th percentiles of that local distribution and estimate the lower-limit on the kick velocity
\begin{equation}
    v_{\rm kick,min}=\max(0,V_{\rm pec,local}-V_{68}).
\end{equation}
Thus, a Gaia NS inside the local 68th-percentile velocity contour has $v_{\rm kick,min}=0$, while a system outside that contour is assigned only the excess above the local stellar velocity. 
This is not a direct natal kick measurement: converting it into a natal kick would require the birth location, Galactic orbit, binary center-of-mass recoil, and pre-supernova orbit.

We also compute the alternative statistic used by \citet{Nagarajan2025a}. 
Instead of comparing the total velocity to the local distribution, this statistic measures the distance from the local median. 
Defining $V_\perp=(V_R^2+V_Z^2)^{1/2}$, we compute
\begin{equation}
    V'_{\rm pec,local}
    =
    \left[
    \left(V_\perp-\widetilde{V}_{\perp,{\rm local}}\right)^2+
    \left(V_\phi-\widetilde{V}_{\phi,{\rm local}}\right)^2
    \right]^{1/2},
\end{equation}
where $\widetilde{V}_{\perp,{\rm local}}$ and $\widetilde{V}_{\phi,{\rm local}}$ are the medians of the local control sample. 
In Figure~\ref{fig:toomre}, this corresponds roughly to the distance between the red star and the gold star. 
We then define
\begin{equation}
    v'_{\rm kick,min}=\max(0,V'_{\rm pec,local}-V'_{68}),
\end{equation}
where $V'_{68}$ is the 68th percentile of the control-star $V'_{\rm pec,local}$ distribution. 
Table~\ref{tab:gaia_ns_kinematics} reports both statistics for each Gaia NS candidate.

\subsection{Be+NS kinematics table}\label{subapp:toomre_bens}
Table \ref{tab:bens_kinematics} provides the kinematics and references used for the Be NS constraints in Section \ref{subsec:bens_kinematics}.

\vspace{-0.8cm}
\begin{deluxetable*}{llrrrrrl}
\tablecaption{Be+NS systems used for the peculiar-velocity comparison.
\label{tab:bens_kinematics}}
\tablehead{
\colhead{System} &
\colhead{Branch} &
\colhead{$P_{\rm orb}$} &
\colhead{$e$} &
\colhead{$\gamma_{\rm sys}$} &
\colhead{$\varpi/\sigma_\varpi$} &
\colhead{$V_{\rm pec}^{\dagger}$} &
\colhead{Refs.} \\
\colhead{} &
\colhead{} &
\colhead{(d)} &
\colhead{} &
\colhead{($\kms$)} &
\colhead{} &
\colhead{($\kms$)} &
\colhead{}
}
\startdata
\multicolumn{8}{l}{\textit{Published systemic radial velocity}} \\
SGR 0755-2933 & low-$e$ & 59.69 & 0.060 & $58.07\pm0.35$ & 20.5 & $6.4^{+0.7}_{-0.4}$ & [5,6] \\
X Per & low-$e$ & 250.3 & 0.111 & $1.0\pm0.9$ & 43.4 & $12.1^{+0.2}_{-0.2}$ & [3,4,27] \\
4U 2206+543 & high-$e$ & 9.558 & 0.300 & $-54.5\pm1.0$ & 22.4 & $22.2^{+0.8}_{-0.8}$ & [25,26] \\
1A 0535+262 & high-$e$ & 110.3 & 0.470 & $-30.0\pm4.0$ & 22.6 & $41.9^{+3.6}_{-3.9}$ & [10,11,12,28] \\
PSR B1259-63 & high-$e$ & 1236.7 & 0.870 & $0.0\pm1.0$ & 33.3 & $24.3^{+1.1}_{-1.2}$ & [17,18,29,30] \\
\hline
\multicolumn{8}{l}{\textit{No published systemic radial velocity}} \\
Swift J0243.6+6124 & low-$e$ & 28.3 & 0.092 & \nodata & 16.0 & $10.6^{+5.5}_{-3.5}$ & [1,2] \\
4U 0115+634 & high-$e$ & 24.317 & 0.339 & \nodata & 8.3 & $23.1^{+13.7}_{-8.2}$ & [7,8] \\
SAX J2103.5+4545 & high-$e$ & 12.665 & 0.406 & \nodata & 10.1 & $37.4^{+22.0}_{-13.3}$ & [23,24] \\
EXO 2030+375 & high-$e$ & 46.022 & 0.410 & \nodata & 4.0 & $11.8^{+7.3}_{-4.4}$ & [21,22] \\
V 0332+53 & high-$e$ & 36.5 & 0.417 & \nodata & 6.7 & $21.9^{+11.0}_{-7.6}$ & [8,9] \\
GX 304-1 & high-$e$ & 132.189 & 0.524 & \nodata & 38.3 & $25.8^{+12.9}_{-8.6}$ & [15,16] \\
GRO J1008-57 & high-$e$ & 247.8 & 0.680 & \nodata & 18.4 & $22.1^{+11.6}_{-7.4}$ & [13,14] \\
PSR J2032+4127 & high-$e$ & $1.7\times10^4$ & 0.961 & \nodata & 36.4 & $42.1^{+21.1}_{-14.4}$ & [19,20] 
\enddata
\tablecomments{
All systems have Gaia DR3 astrometry and
$\varpi/\sigma_\varpi\geq4$.  The quoted intervals are 16th--84th
percentiles.  For systems with a published center-of-mass radial velocity,
$V_{\rm pec}^{\dagger}$ is the direct 3D peculiar speed.  For
the remaining systems without RVs, we marginalize assuming isotropy:
$V_{\rm pec}^{\dagger}=(4/\pi)V_{\rm pec}$.}
\textit{References:}
[1] \citet{doroshenko_orbit_2018};
[2] \citet{reig_optical_2020};
[3] \citet{delgado-marti_orbit_2001};
[4] \citet{zamanov_spectral_2019};
[5] \citet{doroshenko_sgr_2021};
[6] \citet{richardson_high-mass_2023};
[7] \citet{negueruela_bex-ray_2001};
[8] \citet{raichur_apsidal_2010};
[9] \citet{negueruela_bex-ray_1999};
[10] \citet{finger_quasi-periodic_1996};
[11] \citet{wang_constraints_1998};
[12] \citet{okazaki_natural_2001};
[13] \citet{coe_discovery_1994};
[14] \citet{coe_now_2007};
[15] \citet{parkes_shell_1980};
[16] \citet{sugizaki_luminosity_2015};
[17] \citet{negueruela_astrophysical_2011};
[18] \citet{shannon_kinematics_2014};
[19] \citet{lyne_binary_2015};
[20] \citet{ho_multiwavelength_2017};
[21] \citet{motch_accretion_1991};
[22] \citet{fu_timing_2023};
[23] \citet{baykal_timing_2007};
[24] \citet{reig_discovery_2004};
[25] \citet{blay_multiwavelength_2006};
[26] \citet{stoyanov_orbital_2014};
[27] \citet{grundstrom_joint_2007};
[28] \citet{hutchings_spectroscopic_1984};
[29] \citet{1994MNRAS.268..430J};
[30] \citet{Fortin2022}.
\end{deluxetable*}

\section{High Mass-Loss scenario}\label{app:high_mhe}

\begin{figure}
    \centering
    \includegraphics[width=0.99\linewidth]{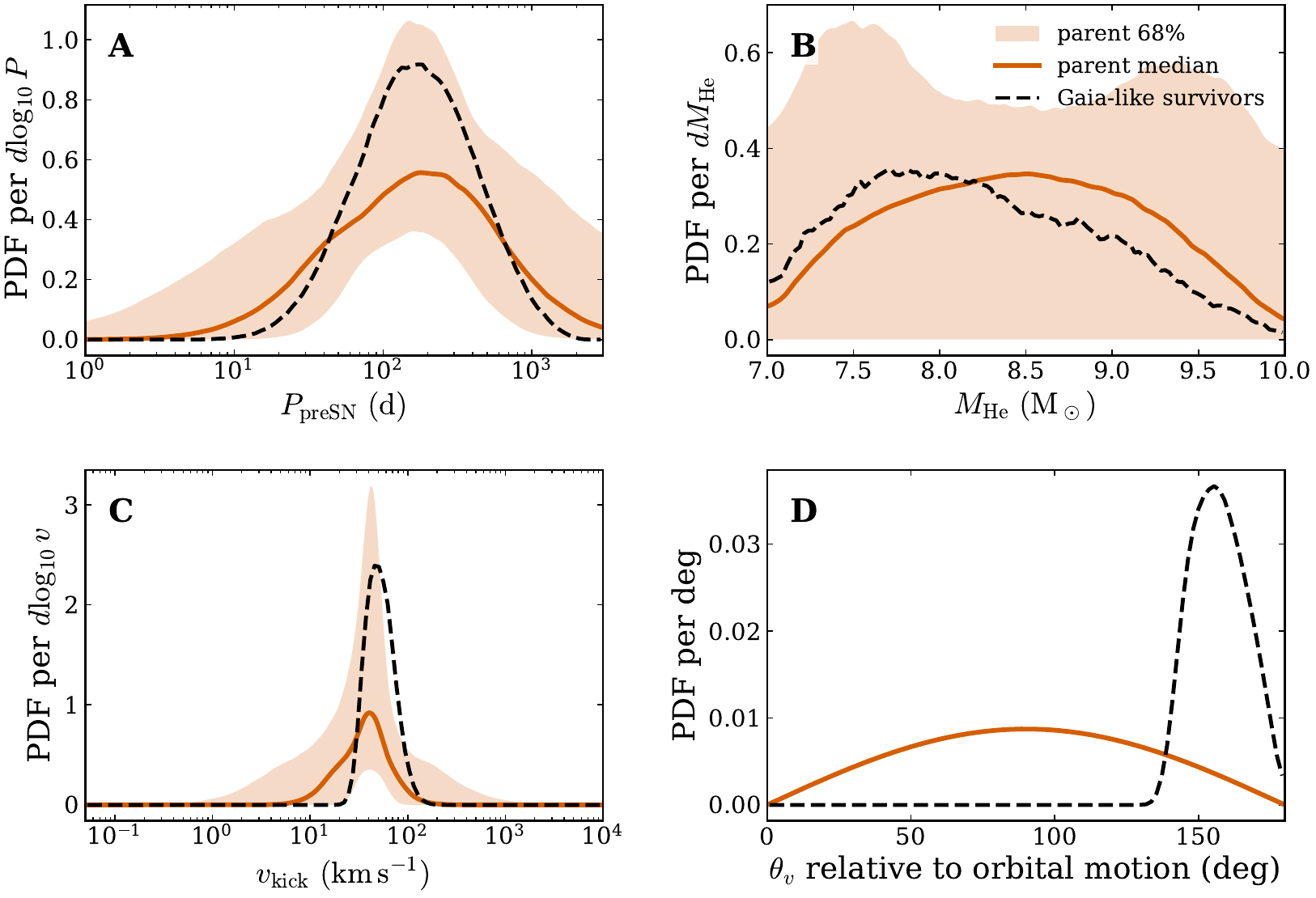}
    \caption{Kick constraints for high-mass progenitors.  The orange curves show the parent distributions before selecting Gaia NS-like survivors, where the shaded regions show the central 68\% posterior range and solid curves show the posterior median. The dashed black curves show the subset of systems that remain bound and land in the Gaia NS period--eccentricity range.  The fit is restricted to $7<M_{\rm He}/\msun<10$, so successful Gaia NS-like survivors require large mass loss and strongly retrograde kicks.}
    \label{fig:high_mhe_stress}
\end{figure}

Allowing a broad progenitor mass distribution tends to favor low $M_{\rm He}$ because less supernova mass loss produces smaller Blaauw kicks and makes binary survival easier (Figure~\ref{fig:survivor_fourpanel_appendix}). Low $M_{\rm He}$ usually implies substantial stripping, as inferred for Be+NS binaries \citep{Valli2025}, but how Gaia NS progenitors could be stripped is unclear. Their moderate eccentricities ($e=0.1$--$0.8$) suggest relatively wide pre-supernova orbits, whereas strong stripping at low mass ratios is usually associated with common-envelope evolution and thus substantial orbital shrinkage to eject the envelope \citep{Webbink1984,Ivanova2013}. An alternative is that the progenitors were only weakly stripped and the binary remained wide but lost more mass in the supernova. We test whether such progenitors can reproduce the Gaia NS orbits.

We repeat the Gaia-only fit (Section~\ref{sec:gaia_only}; Equation~\ref{eq:gaia_likelihood}), restricting progenitor masses to $7<M_{\rm He}/\msun<10$, with $\Delta M=M_{\rm He}-1.4~\msun$. We refit the means and widths of the period, mass, and kick speed distributions, keeping all other assumptions and survivor cuts unchanged. The MCMC uses 40 walkers and 700 steps, discarding the first 250 as burn-in. The final sample contains 52,605 Gaia NS-like survivors from $2\times10^6$ progenitor draws.

The high-mass-loss model matches the Gaia period--eccentricity distributions nearly as well as the fiducial fit ($\Delta\ln{\cal L}=-0.29$), but forms Gaia NSs less efficiently: $\eta_{\it Gaia}=0.022$, compared to $0.083$ in the fiducial stripped model. Figure~\ref{fig:high_mhe_stress} shows the parent and survivor distributions. Successful survivors have $P_{\rm preSN}=174$ d ($56$--$487$ d), $v_{\rm kick}=48.5~\kms$ ($33.9$--$73.3~\kms$), and $\theta_v=154^\circ$ ($143^\circ$--$166^\circ$). Of these, $74\%$ have $\theta_v>150^\circ$, compared to $39\%$ in the fiducial model. Without a natal kick, the mass loss alone gives $e_{\rm Blaauw}=\Delta M/(M_{\rm NS}+M_2)=2.3>1$, unbinding the system. Survival therefore strictly requires a kick that cancels part of the pre-supernova orbital motion.

For the fiducial Gaia NS birth rate ratio $q_{\it Gaia}=5\times10^{-4}$, the required channel fraction is
\begin{equation}
    f_{{\rm ch},{\it Gaia}}=\frac{q_{\it Gaia}}{\eta_{\it Gaia}}
    \approx2\times10^{-2}.
\end{equation}
Namely, $\approx2\%$ of all NS births would need to pass through this configuration to explain the Gaia NS population, or $0.4$--$4.5\%$ for $q_{\it Gaia}=10^{-4}$--$10^{-3}$. This is $\sim4\times$ the fraction required by the fiducial stripped model because fewer systems survive. The Gaia NS abundance therefore does not rule out this channel, but making it dominant requires a non-negligible fraction of NS progenitors to avoid engulfing a solar-type companion in $\sim1$ au orbits.

Our model does not explain how $7$--$10~\msun$ progenitors could survive the RSG or super-AGB phase in these orbits. Observed RSG radii span $R_\star\sim100$--$1000~{\rm \rsun}$ \citep{Yang2020}, which would likely interact with au-scale companions. 
The observed orbits give a direct constraint: the separation at the explosion cannot exceed the final apocenter, so $a_{\rm preSN}\leq a_{\rm final}(1+e_{\rm final})$ for circular progenitors. The observed Gaia NS sample has $2$, $5$, $18$, and $26$ of the $27$ with apocenters below $300$, $500$, $800$, and $1000~\rsun$, respectively. Progenitors with such large radii therefore cannot explain the full sample in this model without interacting with their companions before collapse.

Low metallicity can delay or reduce donor expansion \citep[e.g.,][]{Brunish1982,Baraffe1991,Klencki2020}, and Gaia NSs are significantly overrepresented at low metallicity \citep{ElBadry2024b}. Super-AGB radii remain uncertain and sensitive to mass loss, mixing, and binary interaction \citep[e.g.,][]{Doherty2017}. Models of the maximum progenitor radii and their effects on au-scale companions would help test this scenario.

\section{Constraints on low-mass-ratio binaries from Gaia NSs}\label{app:lowq_progenitor}

The abundance of Gaia NSs in the Galaxy can be used to constrain massive star evolution.  Let $N_{\rm NS}$ be the total number of NSs in the Galaxy and $N_{\rm Gaia~NS}$ be the number of detached Gaia NSs.  In an isolated binary channel,
\begin{equation}
    \frac{N_{\rm Gaia~NS}}{N_{\rm NS}}
    \approx
    f_{\rm bin}\,
    f_{q<0.1,a}\,
    f_{{\rm He+MS},a}\,
    f_{\rm survive}.
    \label{eq:lowq_supply_constraint}
\end{equation}
Here $f_{\rm bin}$ is the massive-star binary fraction, $f_{q<0.1,a}$ is the fraction of massive-star binaries with both a low-mass companion and an initial separation capable of producing an au-scale Gaia NS progenitor, $f_{{\rm He+MS},a}$ is the conditional fraction of those systems that avoid merger and reach an au-scale He-star + MS phase before collapse, and $f_{\rm survive}$ is the fraction that survive NS formation and become Gaia NSs.  We define $q=M_2/M_1$, where $M_2$ is the lower-mass companion and $M_1$ is the NS progenitor. 

\begin{figure}
    \includegraphics[width=0.8\columnwidth]{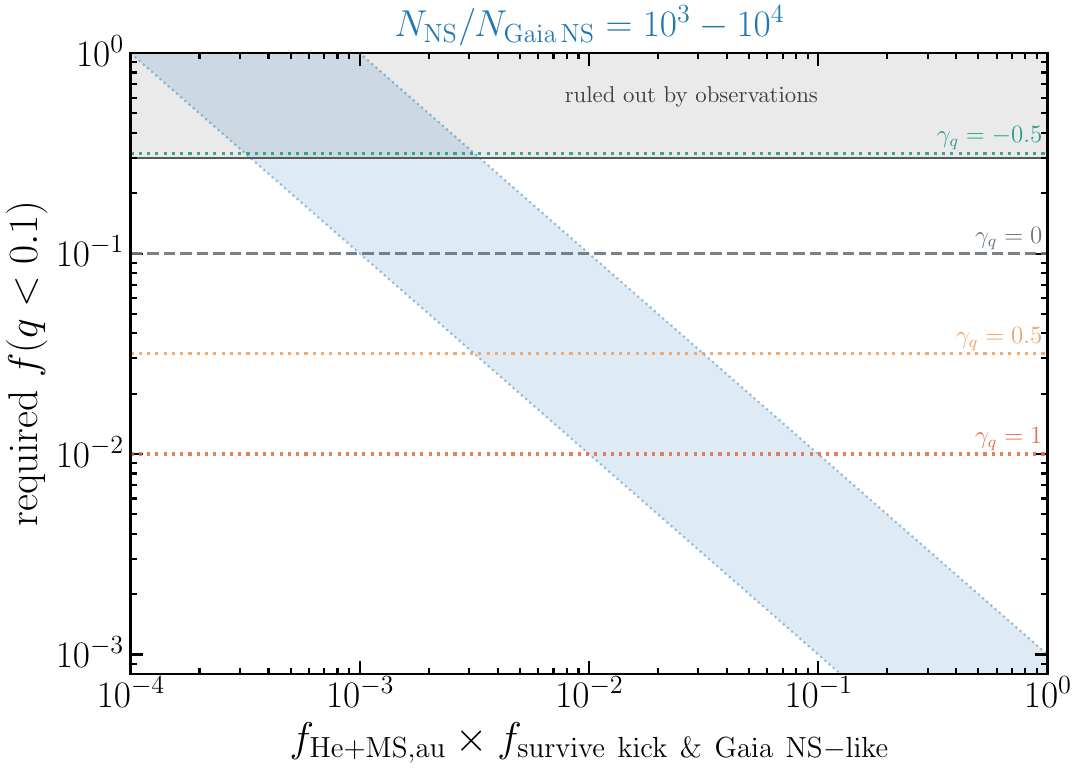}
    \caption{Fraction of low-mass-ratio progenitors required to produce the observed rate of Gaia NSs.  The blue band shows the required $f_{q<0.1,a}$ for $N_{\rm NS}/N_{\rm Gaia~NS}=10^3$--$10{,}000$ in Equation~\ref{eq:lowq_supply_constraint}. The horizontal lines show the fraction of companions with $q<0.1$ for power-law mass-ratio distributions $p(q)\propto q^{\gamma_q}$ over $0.01<q<1$.  The gray region marks $f(q<0.1)>0.3$, where low-mass companions would be more common than expected from the observed massive-star companion population at $q\gtrsim0.1$.}
    \label{fig:lowq_supply_aside}
\end{figure}

\citet{Shariat2026GaiaNSDemographics} find $N_{\rm NS}/N_{\rm Gaia~NS}\sim10^3$--$10^4$, so the product of the four factors in Equation~\ref{eq:lowq_supply_constraint} must be
\begin{equation}
    f_{\rm bin}\,
    f_{q<0.1,a}\,
    f_{{\rm He+MS},a}\,
    f_{\rm survive}
    \approx10^{-4}\text{--}10^{-3}.
    \label{eq:lowq_product_constraint}
\end{equation}

Massive-star multiplicity surveys strongly constrain companions with $q\gtrsim0.1$, but the occurrence of companions with $q<0.1$ is poorly measured \citep{Sana2012,Moe2017,Offner2023}.  For a representative massive-star binary fraction $f_{\rm bin}=0.7$, a population ratio $N_{\rm NS}/N_{\rm Gaia~NS}=5\times10^3$, and a kick survival fraction of $f_{\rm survive}=0.067$, Equation~\ref{eq:lowq_supply_constraint} gives
\begin{equation}
\begin{aligned}
    f_{q<0.1,a}
    &\approx
    \left(5000\,f_{\rm bin}\,f_{{\rm He+MS},a}\,f_{\rm survive}\right)^{-1}
    \nonumber\\
    &=0.043
    \left(\frac{0.7}{f_{\rm bin}}\right)
    \left(\frac{0.1}{f_{{\rm He+MS},a}}\right)
    \left(\frac{0.067}{f_{\rm survive}}\right).
    \label{eq:lowq_supply_scaled}
\end{aligned}
\end{equation}

Namely, if $10\%$ of low-$q$ binaries reach the au-scale He-star + MS phase, the required fraction of initial low-$q$ binaries is $f_{q<0.1,a}\approx4\%$ (Figure~\ref{fig:lowq_supply_aside}).
If only $1\%$ of low-$q$ binaries reach the au-scale He-star + MS phase, the required $f_{q<0.1,a}$ rises to $\sim40\%$, which is likely unfeasible (Figure~\ref{fig:lowq_supply_aside}).  
Thus, for a given natal kick prescription, the abundance of Gaia NSs constrains the occurrence of massive-star binaries with $q<0.1$ companions in au-scale orbits.

\clearpage
\bibliography{refs}
\bibliographystyle{aasjournal}

\end{document}